\documentclass{optica-article}

\journal{opticajournal} 

\articletype{Research Article}

\usepackage{float}
\usepackage{xcolor}
\newif\ifshowrevisions
\ifdefined\markedmanuscript
  \showrevisionstrue
\else
  \showrevisionsfalse
\fi
\newcommand{\rev}[1]{\ifshowrevisions\textcolor{red}{#1}\else#1\fi}

\begin{document}

\title{Perturbation-Based Reconstruction of Depth-Resolved OCT Attenuation without Requiring a Fully Decayed Signal Tail}

\author{Jinglun Yu,\authormark{1,*} Yaning Wang,\authormark{1} Shuwen Wei,\authormark{1} and Jin U. Kang\authormark{1}}

\address{\authormark{1}Department of Electrical and Computer Engineering, Johns Hopkins University, 3400 N. Charles St., Baltimore, MD 21218, USA\\}

\email{\authormark{*}jyu146@jhu.edu} 


\begin{abstract*} 
\rev{Depth-resolved attenuation in optical coherence tomography (OCT) can provide quantitative
optical contrast beyond structural imaging.} However, conventional reconstruction
methods can be biased when
the OCT signal does not fully decay within the imaging range or when the distal signal
tail is dominated by noise or \rev{residual reflections from distal sample or support
interfaces}. \rev{Here, we present a perturbation-based Beer--Lambert framework that
combines a zeroth-order exponential estimate with a first-order correction, reducing
reliance on the unreliable distal signal tail below the current depth.}
Numerical A-line simulations, Monte Carlo modeling, multilayer silicone--dye--TiO$_2$
phantoms, and \emph{ex vivo} bovine retina experiments were used to evaluate the method
under finite-depth and heterogeneous imaging conditions. \rev{The numerical A-line tests
showed reduced depth-dependent error against prescribed attenuation profiles; the
Monte Carlo comparisons showed less distal drift relative to model-defined
benchmarks, while the phantom profiles were compared with separately acquired
single-layer OCT calibration values.} \rev{In a Monte Carlo
component test that withheld vessel-mask and vessel-boundary information from
reconstruction and used that information only for post hoc evaluation,} the first-order
correction recovered \rev{61\%} of the
segmentation-informed target vessel contrast and reduced vessel ROI mean absolute error by
\rev{65\%} relative to the fixed zeroth-order baseline. These results suggest that
perturbation-based reconstruction may provide a practical approach for depth-resolved
reconstruction of effective OCT attenuation when the fully decayed A-line condition is not
satisfied.

\end{abstract*}

\section{Introduction}

Optical coherence tomography (OCT) provides depth-resolved structural imaging of biological tissues and is widely used in ophthalmology. More recently, it has been investigated as an imaging modality for intraoperative robotic surgical guidance \cite{Wang2026AutonomousOCTGuided,Lee2021CPOCTSubretinal,Wang2024AutonomousAnastomosis}. However, OCT intensity is determined not only by tissue morphology, but also by scattering, absorption, focusing, and system-dependent sensitivity. As a result, structural OCT images alone often provide limited quantitative information about tissue optical properties \cite{Wang2022BackscatteringReconstruction}. Estimating the depth-resolved attenuation coefficient from OCT signals has therefore become an important approach for extracting quantitative contrast beyond morphology \cite{Wang2023StationaryIteration,Gong2020AttenuationReview}.

Early OCT attenuation measurements often treated the sample as homogeneous and estimated a single attenuation coefficient by fitting the OCT A-line to an exponential decay. This approach is simple, but it averages the signal over depth and is therefore not well suited for layered or spatially heterogeneous tissues. To obtain local attenuation estimates, depth-resolved methods were later developed under the single-scattering assumption, including discrete formulations adapted from ultrasound imaging \cite{Vermeer2014DepthResolved}. These approaches have been applied to skin, vascular imaging, and atherosclerotic plaque characterization \cite{Ghosh2021SkinWound,Xu2008PlaqueBackscatter,VanSoest2010AtheroscleroticAttenuation}.

A remaining difficulty is that the measured OCT signal depends not only on attenuation but also on the backscattering fraction $R(z)$. This term can vary with scatterer size, refractive-index contrast, and numerical aperture (NA), so treating it as constant can introduce systematic bias in attenuation estimation \cite{Almasian2015BackscatterValidation,Liu2017BackscatterTerm,Cannon2021LayerBased}. Layer-resolved or depth-dependent descriptions of $R(z)$ have therefore been explored to improve reconstruction in heterogeneous tissues \cite{Wang2023StationaryIteration,Cannon2021LayerBased,Zhang2025RetinalLayerBased}. Analytical work has examined the precision and accuracy limits of exponential-fitting and depth-resolved attenuation estimates, including Cramér--Rao bounds \cite{Neubrand2023AccuracyPrecision}; optimized depth-resolved estimators have also been reported \cite{Liu2019OptimizedDepthResolved}. Together, these studies show that attenuation estimation is affected not only by noise, but also by assumptions about signal formation, backscattering, and the distal boundary condition of the A-line.

\rev{Recent studies have further improved quantitative OCT analysis by modeling noise and speckle or suppressing multiple-scattering contributions \cite{Seesan2024ScattererDensityNoise,Zhu2024BscanMultiFocusAveraging}.} These approaches improve robustness under nonideal imaging conditions, but a separate practical limitation remains: many depth-resolved attenuation methods use signal information below the depth being estimated. This makes the reconstruction sensitive to the distal part of the A-line. In ideal cases, the OCT signal decays sufficiently within the imaging range and the remaining tail can be treated in a controlled way. In finite-depth measurements, however, the signal may be truncated before full decay, or the distal tail may be dominated by background noise, residual reflections, and multiply scattered photons. Under these conditions, errors in the deeper signal can propagate upward and produce underestimation in superficial regions or overestimation near the bottom of the imaging range. Noise compensation and signal-weighting strategies can reduce part of this bias \cite{Smith2015AutomatedAttenuation,Li2020RobustAttenuation}, but they still depend on assumptions about the distal signal tail. System-dependent effects, including the point-spread function, sensitivity fall-off, and focus-related variations, can further alter the measured tail signal \cite{VanLeeuwen2003AxialPSF,Dwork2016ConfocalFunction}. In deeper or highly scattering tissues, multiple scattering can also change the apparent attenuation and reduce the reliability of single-scattering reconstructions \cite{Almasian2015BackscatterValidation,Wang2023StationaryIteration}.

In this work, we formulated attenuation profile reconstruction as a perturbation problem around a zeroth-order exponential decay. The measured OCT signal is then used to estimate a baseline attenuation term and a first-order correction, so that the attenuation at a given depth is reconstructed primarily from the integrated signal between the upper boundary of the \rev{local reconstruction interval} and the current depth, rather than from the unreliable distal tail below that depth. This formulation is intended for finite-depth OCT measurements in which the distal A-line may be truncated or contaminated by noise, residual reflections, or multiple-scattering contributions. We evaluated the method using numerical A-line simulations, Monte Carlo models, two-layer silicone--dye--TiO$_2$ phantoms, and \emph{ex vivo} bovine retina. These comparisons were used to determine whether the perturbation-based estimate can reduce depth-dependent bias and improve layer-wise attenuation recovery when the fully decayed A-line condition is not satisfied.

\section{Method}

\subsection{Theory}
Following the Beer–Lambert law, the attenuation of a coherent optical beam propagating in a scattering medium can be described by an exponential decay with respect to depth. 
In our swept-source OCT (SS-OCT) system, the remaining sample-arm irradiance (power per unit area) $L(z)$ associated with round-trip propagation to penetration depth $z$ decreases according to \cite{Wang2023StationaryIteration,Vermeer2014DepthResolved,Liu2017BackscatterTerm}
\begin{equation}
\label{eq:beer_lambert}
L(z) = L_0 \exp\!\left(-2 \int_{0}^{z} \mu(u)\, du \right),
\end{equation}
where $L_0$ denotes the incident irradiance and $\mu(z)$ is the depth-dependent attenuation coefficient. 
The factor of 2 accounts for the double-pass propagation of light in OCT, corresponding to forward and backward travel through the tissue.
Taking the spatial derivative yields the equivalent first-order differential form
\begin{equation}
\label{eq:differential_form}
\frac{dL(z)}{dz} + 2\mu(z)L(z) = 0.
\end{equation}
To obtain a first-order approximation of $\mu(z)$, we introduce a regular perturbation expansion in $\varepsilon$. Within each \rev{local reconstruction interval}, the zeroth-order attenuation $\mu_{0}$ is treated as constant, while $\mu_{1}(z)$ describes depth-dependent deviations from the leading exponential decay:
\begin{equation}
\begin{cases}
\mu(z) = \mu_{0} + \varepsilon \mu_{1}(z) + \varepsilon^{2} \mu_{2}(z) + \cdots, \\
L(z) = L_{0}(z) + \varepsilon L_{1}(z) + \varepsilon^{2} L_{2}(z) + \cdots.
\end{cases}
\end{equation}
Substituting these expansions into Eq.~\eqref{eq:differential_form} gives
\begin{equation}
\frac{d}{dz}\!\left(L_{0} + \varepsilon L_{1} + \cdots \right)
+ 2\left(\mu_{0} + \varepsilon \mu_{1} + \cdots \right)
\left(L_{0} + \varepsilon L_{1} + \cdots \right)
= 0.
\end{equation}

Collecting terms with identical powers of $\varepsilon$ leads to a hierarchy of linear ordinary differential equations:
\begin{equation}
\left\{
\begin{aligned}
\mathcal{O}(1):\quad 
& \frac{dL_{0}(z)}{dz} + 2\mu_{0}L_{0}(z) = 0, \\
\mathcal{O}(\varepsilon):\quad 
& \frac{dL_{1}(z)}{dz} + 2\mu_{0}L_{1}(z) + 2\mu_{1}(z)L_{0}(z) = 0, \\
\mathcal{O}(\varepsilon^{2}):\quad 
& \frac{dL_{2}(z)}{dz} + 2\mu_{0}L_{2}(z) + 2\mu_{1}(z)L_{1}(z) + 2\mu_{2}(z)L_{0}(z) = 0.
\end{aligned}
\right.
\end{equation}

The leading-order term $\mathcal{O}(1)$ represents the dominant attenuation behavior, while higher-order terms provide successive corrections. 
Solving the leading-order equation yields
\begin{equation}
L_{0}(z) = L_{0} e^{-2\mu_{0} z}.
\end{equation}

For the first-order correction, we assume a solution of the form
\begin{equation}
L_{1}(z) = C(z)e^{-2\mu_{0} z},
\end{equation}
where $C(z)$ is an auxiliary function to be determined. 
We impose the first-order boundary condition $L_{1}(0)=0$, equivalently
$C(0)=0$, so that the incident irradiance at the interval entrance is represented
entirely by the leading-order term.
Substituting this into the $\mathcal{O}(\varepsilon)$ equation and utilizing the leading-order solution gives
\begin{equation}
C(z) = -2L_{0}\int_{0}^{z}\mu_{1}(u)\,du.
\end{equation}

Neglecting terms of order $\mathcal{O}(\varepsilon^{2})$ and higher and setting the
bookkeeping parameter $\varepsilon$ to unity after truncation gives the first-order approximation
\begin{equation}
\label{eq:first_order_lz}
L(z) \approx L_{0}(z) + L_{1}(z)
= L_{0}\left(1 - 2\int_{0}^{z}\mu_{1}(u)\,du \right)
e^{-2\mu_{0} z}.
\end{equation}

The attenuation coefficient is then written as the sum of its zeroth- and first-order terms,
\begin{equation}
\mu(z) \approx \mu_{0} + \mu_{1}(z).
\end{equation}

The sample-generated OCT signal intensity $I_{\mathrm{s}}(z)$, before multiplication by the system response, is modeled as the backscattered contribution
\begin{equation}
\label{eq:oct_signal_relation}
I_{\mathrm{s}}(z) = -R(z)\frac{dL(z)}{dz},
\end{equation}
where $R(z)$ denotes the physical backscattered fraction.

Substituting the first-order approximation of $L(z)$ (Eq.~\eqref{eq:first_order_lz}) into the above relation yields
\begin{equation}
\label{eq:intensity_first_order}
I_{\mathrm{s}}(z) = 2L_{0}R(z)
\left[
\mu_{1}(z) + \mu_{0}
- 2\mu_{0}\int_{0}^{z}\mu_{1}(u)\,du
\right]
e^{-2\mu_{0}z}.
\end{equation}

To apply the reconstruction piecewise, we assume that the backscattered fraction is constant within each local depth interval, i.e.,
\[
R(z) = R_{ab}, \quad z \in [z_a, z_b],
\]
and introduce the local depth coordinate
\[
\xi=z-z_a,\qquad \xi\in[0,z_b-z_a].
\]
The corrected local linear OCT intensity and first-order attenuation correction are denoted by
$I_{ab}(\xi)=C_{ab}I_{\mathrm{s}}(z_a+\xi)$ and
$\mu_{1,ab}(\xi)=\mu_1(z_a+\xi)$, respectively, where $C_{ab}$
\rev{collects only residual multiplicative system scaling after background subtraction
and confocal-response correction and is treated as constant within the local reconstruction
interval}. Within the interval,
the zeroth-order attenuation $\mu_{0,ab}$ is treated as constant, and
$L_{0,ab}$ denotes the leading-order irradiance at the upper boundary
$z_a$. The corresponding effective local amplitude parameter is
\[
A_{ab} \equiv 2C_{ab}L_{0,ab}R_{ab}.
\]
Thus, $A_{ab}$ absorbs both the local backscattering scale and the
\rev{remaining depth-independent multiplicative system scale}.
Applying Eq.~\eqref{eq:intensity_first_order} in this local coordinate gives the following first-order linear ordinary differential equation:
\begin{equation}
\label{eq:mu1_ode}
\frac{d\mu_{1,ab}(\xi)}{d\xi}
- 2\mu_{0,ab}\mu_{1,ab}(\xi)
=
\frac{1}{A_{ab}}
\left[
I_{ab}'(\xi) + 2\mu_{0,ab}I_{ab}(\xi)
\right]
e^{2\mu_{0,ab}\xi}.
\end{equation}

Evaluating the local form of Eq.~\eqref{eq:intensity_first_order} at the upper boundary of the interval gives
\begin{equation}
\mu_{1,ab}(0)
=
\frac{I_{ab}(0)}{A_{ab}}
-
\mu_{0,ab}.
\end{equation}

The general solution of the above linear ODE can be written as
\begin{equation}
\mu_{1,ab}(\xi) = D_{ab}(\xi)e^{2\mu_{0,ab}\xi},
\end{equation}
where $D_{ab}(\xi)$ is an auxiliary function to be determined.

Applying the boundary condition gives
\begin{equation}
D_{ab}(0) = \frac{I_{ab}(0)}{A_{ab}} - \mu_{0,ab}.
\end{equation}

Substituting into Eq.~\eqref{eq:mu1_ode} and integrating yields
\begin{equation}
D_{ab}(\xi)
=
\int_{0}^{\xi}
\frac{1}{A_{ab}}
\left[
I_{ab}'(s) + 2\mu_{0,ab} I_{ab}(s)
\right] ds
+
\frac{I_{ab}(0)}{A_{ab}}
-
\mu_{0,ab}.
\end{equation}

The first-order correction term is therefore
\begin{equation}
\mu_{1,ab}(\xi)
=
\left[
\frac{I_{ab}(\xi)}{A_{ab}}
+
\frac{2\mu_{0,ab}}{A_{ab}}
\int_{0}^{\xi} I_{ab}(s)\,ds
-
\mu_{0,ab}
\right]
e^{2\mu_{0,ab}\xi}.
\end{equation}

The estimated attenuation coefficient in the interval is then
\begin{equation}
\label{eq:reconstructed_mu}
\hat{\mu}(z)
=
\hat{\mu}_{0,ab} + \hat{\mu}_{1,ab}(z-z_a),
\qquad z\in[z_a,z_b],
\end{equation}
where $\hat{\mu}_{1,ab}$ is obtained from the first-order solution after substituting the fitted parameters. The local zeroth-order attenuation $\mu_{0,ab}$ and effective amplitude $A_{ab}$ are estimated via nonlinear least-squares fitting.
\rev{The samples retained for this fit need not occupy the entire reconstruction interval.
We denote their support by
$[z_{nf0},z_{nf1}]\subseteq[z_a,z_b]$ and define
$\mathcal{K}_{ab}=\{k\,|\,z_a+k\Delta z\in[z_{nf0},z_{nf1}]\}$.}
\begin{equation}
\label{eq:zeroth_order_fit}
(\hat{\mu}_{0,ab}, \hat{A}_{ab})
=
\underset{\mu_{0,ab},\,A_{ab}}{\arg\min}
\rev{\sum_{k\in\mathcal{K}_{ab}}}
\left[
I_{ab,k}
- A_{ab}\mu_{0,ab}e^{-2\mu_{0,ab}\xi_k}
\right]^2,
\end{equation}
where $I_{ab,k}=I_{ab}(\xi_k)$ and $\xi_k=k\Delta z$.
\rev{The retained fitting-support size is
$N_{ab}^{\mathrm{fit}}=|\mathcal{K}_{ab}|$. The fitted pair is applied throughout
$[z_a,z_b]$, while the boundary condition and first-order cumulative integral remain
referenced to $z_a$.}

\rev{Although two distinct samples are sufficient to identify the two parameters in
Eq.~\eqref{eq:zeroth_order_fit} in the noise-free algebraic sense, this is not a useful
stability criterion for measured OCT data. We therefore required the retained fitting
support to contain
\(N_{ab}^{\mathrm{fit}}\geq\max[10,1+\lceil3\delta z_{\mathrm{ax}}/\Delta z\rceil]\) samples, where
\(\delta z_{\mathrm{ax}}\) is the axial resolution in the medium. The additional
sample accounts for the first-to-last physical span of
\((N_{ab}^{\mathrm{fit}}-1)\Delta z\). Thus, every fitted support contains at least
10 samples and spans at least three axial-resolution elements. If
interface exclusion leaves a thinner interval, that region is not assigned an
independently fitted pair \((\hat{\mu}_{0,ab},\hat{A}_{ab})\); it must instead be merged
with an adjacent optically similar interval or supplied with an independently calibrated
baseline. This is a practical fitting-support criterion rather than a lower bound imposed
by the perturbation equations themselves.}

Equation~\eqref{eq:reconstructed_mu} is the reconstruction formula used in this work.
For each interval, the fitting coordinate, boundary condition, and cumulative integral are reset at $\xi=0$, consistent with the piecewise implementation. The integral was evaluated as a cumulative sum on the uniformly sampled axial grid. The zeroth-order term is obtained by the curve fitting in Eq.~\eqref{eq:zeroth_order_fit}.

\rev{The 101- and 51-pixel lateral moving averages (1.01 and 0.51~mm at $10~\mu$m
per A-line) were empirical smoothing settings for displaying layer-wise $\hat{\mu}_0$
component maps, not reconstruction supports. The revised phantom $\hat{\mu}_0$ panels
retain the 101-pixel display smoothing, whereas the earlier 51-pixel smoothing has been
removed from the bovine panels. Neither window is used to generate the final
$\hat{\mu}_0+\hat{\mu}_1$ maps, representative A-line profiles, or quantitative metrics.
The first-order correction in each final map uses corrected linear intensity from the
corresponding A-line. During map reconstruction, segmented boundary traces were
stabilized over 21 neighboring A-lines for the phantoms and 11 for the bovine retina to
reduce boundary jitter.}

For comparison, the conventional depth-resolved attenuation was calculated from the same
sampled OCT intensity according to Vermeer \emph{et al.}~\cite{Vermeer2014DepthResolved}:
\begin{equation}
\label{eq:conventional_dr}
\hat{\mu}_{\mathrm{DR}}(z_i)
=
\frac{I(z_i)}
{2\Delta z\sum_{j=i}^{N}I(z_j)},
\end{equation}
where $\Delta z$ is the axial sampling interval and $N$ is the final sampled depth.

\subsection{Implications of Multiple Scattering}

Multiple scattering can affect attenuation estimates in OCT, particularly in tissues with
strong forward-scattering properties \cite{Faber2004Attenuation}. As imaging depth
increases, multiply scattered photons can contribute to the detected signal and alter the
distal OCT signal relative to the single-scattering exponential decay model. This effect
changes the apparent signal decay and complicates the interpretation of attenuation
estimates in heterogeneous tissues.

This effect can be theoretically interpreted using the Extended Huygens--Fresnel (EHF)
model \cite{Gong2020AttenuationReview}. The model assumes the paraxial approximation
($\sin\theta \approx \theta$), where $\theta$ denotes the scattering angle between the
incident and scattered wave vectors. In this formulation, the mean OCT signal consists
of three components: the singly backscattered field, the multiply forward-scattered field,
and the coherent cross-term between these two contributions. The mean squared OCT
amplitude, equivalent to the OCT intensity, can be written as
\begin{equation}
\label{eq:ehf_intensity}
\langle A^2(z) \rangle \propto
\frac{1}{w_H^2(z)}
\left[
e^{-2\mu_s z}
+
\frac{
4e^{-\mu_s z}\left(1-e^{-\mu_s z}\right)
}{
1+\left(\frac{w_S(z)}{w_H(z)}\right)^2
}
+
\left(1-e^{-\mu_s z}\right)^2
\left(\frac{w_H(z)}{w_S(z)}\right)^2
\right],
\end{equation}
where $w_H(z)$ represents the local beam waist in the absence of forward scattering and
$w_S(z)$ denotes the effective beam waist in the presence of forward multiple scattering.
The parameter $w_0$ is the beam waist at the focal plane in air, and the root-mean-square
scattering angle $\theta_{\mathrm{RMS}}$ is related to the scattering anisotropy through
$\theta_{\mathrm{RMS}} \approx \sqrt{2(1-g)}$. The EHF model extends the
single-scattering description by including multiply forward-scattered and coherent
cross-term contributions. The Beer--Lambert single-scattering form is recovered when these
additional contributions are negligible. Therefore, the EHF formulation provides theoretical
context for understanding how forward multiple scattering can alter the distal OCT signal
and violate the ideal single-scattering exponential decay assumption.

In conventional tail-integral-based attenuation reconstruction, the attenuation estimate
at a given depth depends on the signal below that depth. Therefore, distal signal changes
caused by multiple scattering can propagate into shallower attenuation estimates and
introduce depth-dependent bias. This effect is especially relevant in layered tissues, such
as the retina, where deeper scattering regions can influence the reconstructed attenuation
in more superficial layers.

The perturbation-based reconstruction does not eliminate multiple scattering and should
not be interpreted as recovering the intrinsic microscopic scattering coefficient in
strongly multiple-scattering regions. Instead, by reducing reliance on the distal
signal tail, it limits the propagation of distal signal errors into more superficial
attenuation estimates. Accordingly, the reconstructed attenuation should be interpreted
as an effective OCT attenuation parameter under strong multiple-scattering conditions.
This distinction also clarifies the complementary roles of the two modeling frameworks
used in this study: the EHF formulation provides theoretical context for how forward
multiple scattering alters the OCT signal, whereas Monte Carlo simulations provide the
numerical evaluation of its effects on attenuation reconstruction.

The retinal layer parameters used in the subsequent Monte Carlo simulation are summarized in
Table~1 \cite{Hammer1995OcularFundus,Gerakis2008MonteCarloOCT,Jacques2013OpticalProperties,Chen2015RetinalOximetryMC}.

\begin{table}[H]
\caption{Retinal Layer Thickness and Optical Properties}
\label{tab:retina_optical}
\centering
\small
\renewcommand{\arraystretch}{1.15}
\begin{tabular}{cccccc}
\hline
Layer & $d$ ($\mu$m) & $n$ & $\mu_a$ (cm$^{-1}$) & $\mu_s$ (cm$^{-1}$) & $g$ \\
\hline
Retina & 200 & 1.47 & 1 & 224 & 0.97 \\
RPE & 10  & 1.47 & 385 & 1469 & 0.84 \\
Choroid & 250 & 1.47 & 27 & 484 & 0.94 \\
Sclera & 700 & 1.47 & 0.4 & 614 & 0.90 \\
Blood Vessel & 200 & 1.47 & 2 & 140 & 0.84 \\
\hline
\end{tabular}
\end{table}

\subsection{Numerical Simulation}
\subsubsection{OCT A-line Numerical Imaging}
The first numerical simulation was designed to isolate the effect of finite imaging depth. OCT A-lines
were generated under the ideal single-scattering Beer--Lambert model, and the imaging range
was intentionally truncated before the signal fully decayed within the imaging window. This
setting directly tests the failure mode of the conventional tail-integral reconstruction methods, in
which the attenuation estimate depends on the signal information below the depth of interest. It
therefore allowed us to evaluate whether each method could recover the prescribed
attenuation profile without relying on the missing distal signal tail. OCT depth profiles
$I(z)$ were generated based on Eq.~\eqref{eq:beer_lambert} and Eq.~\eqref{eq:oct_signal_relation} by specifying the attenuation coefficient
$\mu(z)$ and backscattering fraction $R(z)$. \rev{Both profiles used a depth-independent backscattering fraction $R(z)=R_0$, so the normalized value $R(z)/R_0=1$ throughout. Because each A-line was normalized to unit peak, no absolute physical value of $R_0$ was assigned.} The first profile contained step changes, including
$\mu=0.2~\mathrm{mm}^{-1}$ at $z=0.3$~mm and $\mu=1~\mathrm{mm}^{-1}$ at
$z=1.4$~mm, whereas the second followed a smoothly varying sinusoidal profile.
\rev{The simulated sample surface was positioned at the zero-delay plane.} The combined incident irradiance and constant
system factor was set to arbitrary values of $L_0\beta=1.67$ for \rev{the step-like case} and
$L_0\beta=1$~A.U. for \rev{the smoothly varying case}. \rev{The prescribed attenuation
values provided illustrative numerical profiles.} Each simulated profile contained 1000 axial samples
with an axial sampling interval of $2~\mu$m, corresponding to a total imaging depth of
$2~\mathrm{mm}$. The profiles were analyzed over three principal depth intervals of 600,
600, and 800~$\mu$m. \rev{Each clean linear-intensity profile was normalized to unit peak
before adding zero-mean white Gaussian noise. The nominal unit-reference SNR was defined by
$\sigma_n=10^{-\mathrm{SNR}_{\mathrm{nom}}/20}$, where $\sigma_n$ is the noise standard
deviation relative to the unit peak. The step-like and smoothly varying cases used
$\mathrm{SNR}_{\mathrm{nom}}=40$ and 60~dB, respectively, corresponding to
$\sigma_n=0.01$ and 0.001. When calculated instead from the mean-square power of each
clean profile, the resulting SNRs were 33.1 and 51.0~dB, respectively. These were separate
numerical stress conditions rather than a
matched-noise comparison: the lower-SNR step case tested boundary recovery under stronger
noise, whereas the higher-SNR smooth case allowed continuous-profile tracking to remain
visually distinguishable from stochastic fluctuations.} No separate constant noise-floor
offset was imposed. Under this
configuration, the simulated A-line represents an idealized OCT signal that excludes
practical system effects, including multiple scattering, confocal focusing, and sensitivity
roll-off. Multiple-scattering effects were evaluated separately using Monte Carlo
simulations, as described below.

\rev{An additional finite-tail benchmark compared the proposed reconstruction with the
stationary-iteration method (SIM) and optimized depth-resolved estimation (ODRE). A common
piecewise attenuation profile was simulated over three regions with layer-dependent
backscattering fractions. Two matched 40-dB-SNR A-lines were evaluated: a clean-tail case
and a case containing a controlled Gaussian residual reflection centered at 1.90~mm. All
methods received the same linear-intensity A-line, axial sampling, layer boundaries when
required, and evaluation mask. All fitting windows and algorithm parameters were fixed
before generating the residual-reflection case and were unchanged between the two cases.
The two cases used an identical noise realization, and only the residual reflection differed.
SIM was independently implemented from the published attenuation initialization,
backscattering update, and joint iteration equations of Wang et
al.~\cite{Wang2023StationaryIteration}. Samples 711--980 were used for the
distal exponential fit, samples 981--1000 were replaced by that fit, and the fitted tail was
extrapolated until its residual fraction reached $10^{-10}$, following the noise-floor
extrapolation and attenuation-initialization procedure. The backscattering profile was then
updated from the current attenuation estimate, retaining transitions whose absolute axial
derivative exceeded 4.5 times its mean and replacing each intervening low-variation interval
by its mean. The attenuation and backscattering estimates were jointly updated until their
maximum relative change was at most 1\% or 100 iterations were reached. Because the
constraint factors are application-dependent in the published description, deliberately
broad ratio bounds of 0.1--10 were applied to both the backscattering and attenuation
updates; neither bound was activated in the reported cases. Sensitivity analysis additionally
used derivative multipliers of 4, 4.5, and 5 and ratio bounds of 0.1--10, 0.25--4, and
0.5--2. Following Liu
et al.~\cite{Liu2019OptimizedDepthResolved}, the
final 120 samples used for ODRE were fitted to an exponential-plus-constant model, and the
fitted terminal intensity and attenuation were included in the reconstruction denominator.
Mean absolute error (MAE) was calculated after excluding the first and last 10 samples and
10 samples on either side of each prescribed attenuation or backscattering transition. The
evaluation mask was further divided into the proximal 80\% and distal 20\% of the simulated
depth range.}

\subsubsection{Monte Carlo Simulation}

\rev{The Monte Carlo analysis comprised three linked but distinct tests. The first isolated
the effect of strongly forward scattering in a generic multilayer phantom. The second tested
the complete reconstruction in a spatially heterogeneous retina-like geometry. The third
tested the first-order contribution in that retina-like model while withholding vessel-mask
and vessel-boundary information during reconstruction.}

Unlike the ideal Beer--Lambert A-line simulations, the Monte Carlo model explicitly accounts for photon scattering histories and therefore provides a direct numerical framework for evaluating multiple-scattering-induced deviations from the single-scattering signal model. Light propagation in the multilayer tissue was simulated using a Monte Carlo approach based on the MCML algorithm originally developed by Wang \emph{et al.} \cite{Wang1995MCML,Gerakis2008MonteCarloOCT}. Photon packets were initialized with a weight $w=1$ and progressively attenuated through absorption and scattering events during propagation. Only photons satisfying the OCT detection conditions (NA, detector position, and detector size) were recorded.

Following the convention of Gerakis \emph{et al.} \cite{Gerakis2008MonteCarloOCT}, the OCT signal intensity envelope used for attenuation reconstruction was computed in the optical path-length coordinate $\ell$ as
\begin{equation}
I_{\mathrm{env}}(\ell)
=I_0\sum_i
\sqrt{w_r w_s(\Delta l_i)}
\exp\!\left[
-\left(
\frac{\ell-\Delta l_i}{l_{\mathrm{coh}}}
\right)^2
\right],
\end{equation}
where $I_0$ is a source-dependent constant, $w_r$ is the reference-arm photon weight, $w_s(\Delta l_i)$ is the sample-arm photon weight associated with the $i$th detected photon, $\Delta l_i$ is its optical path-length mismatch relative to the reference arm, and $l_{\mathrm{coh}}$ denotes the coherence length of the source. Here, $\ell$ denotes the round-trip optical path-length coordinate. Before attenuation reconstruction, $\ell$ was converted to the physical penetration depth $z$ according to $\ell(z)=2\int_0^z n(u)\,du$. Throughout the Monte Carlo analysis, the resulting intensity envelope is denoted by $I(z)$ when applying the attenuation reconstruction methods.

\paragraph{\rev{Test 1: multiple-scattering stress test.}}
\rev{This test compared paired three-layer Monte Carlo-derived A-lines under isotropic
($g=0$) and strongly forward-scattering ($g=0.9$) conditions. The paired simulation
inputs each used $7\times10^6$ photons, three $300~\mu\mathrm{m}$ layers,
refractive index $n=1.4$ in the layers and surrounding media, and layer-wise
$\mu_a=\mu_s=(30,50,20)~\mathrm{cm}^{-1}$. The axial and radial bin widths
were both $1~\mu\mathrm{m}$, with 900 axial bins and one radial and angular bin;
only $g$ was changed. Each supplied linear-intensity A-line thus contained
900 samples spanning 0.9~mm, with layer boundaries at 0, 0.3, 0.6, and
0.9~mm. The displayed 420-column B-scans were constructed by placing copies of
each A-line at surface samples
\(k_s(j)=\operatorname{round}[160+25\sin(4\pi j/420)]\),
where \(j=0,\ldots,419\) indexes the displayed columns. The internal boundaries
at offsets of 300 and 600 samples followed the same translation for both
scattering conditions. The columns are not independent lateral Monte Carlo
realizations.}
\rev{For each scattering condition, a layer-wise exponential fit to its own linear
intensity profile defined the condition-matched effective-attenuation reference.
The proposed and conventional depth-resolved methods were applied to the same
condition-specific A-line. Statistics were calculated over the central 80\% of
each layer, excluding 10\% on either side of the interfaces. The $g=0.9$ minus
$g=0$ reference difference measures the change in the effective signal decay;
signed bias and MAE relative to each condition's own fitted reference measure
additional reconstruction error. The $g=0$ comparison is an isotropic-scattering
condition, not an isolated single-scattering component.}

\paragraph{\rev{Test 2: retina-like heterogeneous reconstruction.}}
\rev{This test evaluated whether the complete method preserved layer-wise attenuation contrast
in a curved, heterogeneous geometry containing an embedded vascular structure.}
The anatomically inspired retina-like model was adapted from previous studies and consisted of four tissue layers: retina, retinal pigment epithelium (RPE), choroid, and sclera, with embedded vascular structures. Because all retinal layers were assigned $n=1.47$, the optical-path conversion reduced to $z=\ell/(2n)$.
For the retina-like configuration, the system parameters included a numerical aperture of 0.05 and a coherence length of $5~\mu$m. Two Monte Carlo A-line intensity profiles, representing retinal tissue without and with a vascular contribution, were used as source profiles. Each source profile contained 1000 axial samples with an axial sampling interval of $1~\mu$m, and a total of $5\times10^6$ photon packets was used to generate the underlying Monte Carlo profiles. A synthetic B-scan containing 2000 lateral columns was then constructed by piecewise mapping the source profiles onto the prescribed retinal layer boundaries, with the vascular profile blended within the vessel mask. The constructed B-scan had a lateral sampling interval of $4~\mu$m per pixel, and the central 1400 columns shown in Fig.~\ref{fig:fig3_1} corresponded to a lateral field of view of $5.6~\mathrm{mm}$.

\rev{The original multiplicative Gaussian modulation was extended from a pixel-wise
independent field to a resolution-correlated Gaussian field. Specifically, a real white
Gaussian field $W(z,x)\sim\mathcal{N}(0,1)$ was convolved with a separable Gaussian
kernel $h(z,x)$ and normalized as}
\begin{equation}
\rev{\xi_c(z,x)=\frac{[h*W](z,x)-\mathbb{E}[h*W]}
{\sqrt{\operatorname{Var}[h*W]}},
\qquad \xi_c\sim\mathcal{N}(0,1).}
\label{eq:correlated_gaussian_field}
\end{equation}
\rev{The kernel widths were selected to give the separable covariance}
\begin{equation}
\rev{\operatorname{Cov}(\Delta z,\Delta x)=
\exp\!\left[-4\ln 2\left(
\frac{\Delta z^2}{L_z^2}+\frac{\Delta x^2}{L_x^2}\right)\right],}
\label{eq:correlated_gaussian_covariance}
\end{equation}
\rev{whose axial and lateral FWHMs were $L_z=5.0~\mu$m and $L_x=12.9~\mu$m,
respectively, matching the axial coherence length and lateral resolution. This construction
follows the physical expectation that OCT speckle correlation is set by the imaging
point-spread function \cite{Chong2018SpeckleAutocorrelation}. The perturbed linear intensity
was then defined as}
\begin{equation}
\rev{I_{\mathrm{pert}}(z,x)=\max\!\left\{0,
I_{\mathrm{MC}}(z,x)M(z,x)+b+\sigma_n\eta(z,x)\right\},
\quad M(z,x)=\max\!\left\{0,1+\sigma_m\xi_c(z,x)\right\},}
\label{eq:stochastic_corruption}
\end{equation}
\rev{where $\eta$ is an independent real standard-normal field, $\sigma_m=0.18$,
$b=2\times10^{-4}$, and $\sigma_n=1\times10^{-3}$ in the source-profile intensity scale.
Across eight independent realizations, the measured correlation FWHMs were
$5.030\pm0.005~\mu$m and $13.032\pm0.017~\mu$m. Figures~\ref{fig:fig3_1} and
\ref{fig:vessel_ablation}(a--e) show the realization with the smallest standardized distance
from the ensemble means of the reported contrast and MAE metrics; the metrics in
Fig.~\ref{fig:vessel_ablation}(f) and the reported ablation statistics are the mean and
standard deviation across all eight realizations.}

Confocal focusing and weak sensitivity roll-off were applied only when rendering the simulated OCT intensity profiles and image in Fig.~\ref{fig:fig3_1}(a,b) and were not included in the attenuation-reconstruction input. Throughout the numerical A-line and Monte Carlo analyses, attenuation reconstruction was performed using linear OCT intensity; display transformations, including logarithmic compression, were used only for visualization.

\paragraph{\rev{Test 3: first-order contribution without vessel segmentation.}}
\rev{This component test asked whether the first-order term could restore vessel-associated local
attenuation variation when the zeroth-order baseline contained only remote layer-wise fits and
no vessel location or boundary information.}
\rev{The component test} used the same retina-like B-scan. The vessel mask and vessel boundaries were withheld from parameter estimation and reconstruction. Fixed layer-wise zeroth-order decay and amplitude parameters were obtained as the median fitted values from remote control columns outside the vascular region and were applied within the corresponding retinal layer boundaries. The first-order correction was then calculated independently from the local linear OCT intensity in each A-line. The vessel and surrounding-retina ROIs were defined only after reconstruction for quantitative evaluation. Vessel contrast was calculated as the difference between the median attenuation in the vessel ROI and that in the adjacent surrounding-retina ROI, and the vessel ROI mean absolute error (MAE) was calculated relative to the segmentation-informed fitted target.

\subsection{Phantom preparation and biological samples}

To evaluate the performance of the proposed method, two-layer silicone--elastomer phantoms were
\rev{constructed following established silicone-based optical-phantom fabrication approaches
\cite{Ayers2008SiliconePhantoms,Curatolo2011StructuredPhantom}. TiO$_2$ was used as the
scattering agent, whereas Higgins\textsuperscript{\textregistered} black dye was used as a
broadband absorber to vary the absorption contribution to the OCT signal.} Viscous mixtures consisting of
TiO$_2$ particles, black dye (Higgins\textsuperscript{\textregistered}), and a base silicone
elastomer (Sylgard\textsuperscript{\textregistered} 284, MicroLubrol), together with its curing agent, were
thoroughly mixed to achieve homogeneous optical properties. Phantom~1 contained 0.05~wt\%
TiO$_2$ in both layers and \rev{0~wt\% and 0.5~wt\%} black dye in the upper and lower layers,
respectively. Phantom~2 contained 0.025~wt\% and 0.05~wt\% TiO$_2$ and 0~wt\% and
1.5~wt\% black dye in the corresponding layers. Each mixture was cast and cured into layers
approximately 300--600~$\mu$m thick. After curing, individual layers were imaged and assembled
into the two-layer configurations used for OCT measurements. \rev{The TiO$_2$ and dye
concentrations were selected to produce layer-dependent scattering and absorption contrast;
the nominal compositions were not treated as independent measurements of $\mu_s$ or $\mu_a$.}

\rev{For evaluation in biological tissue,} \emph{ex vivo} bovine retinas were used. Healthy eyes were dissected to
expose the retinal surface in an open-sky configuration, enabling direct OCT imaging without
additional optical interfaces. \rev{No matched \emph{in vivo} scan, histological reference, or
independent assessment of outer-retinal preservation was available for these specimens;
therefore, this experiment was not designed to separate the effects of bovine anatomy,
open-sky preparation, postmortem tissue change, and acquisition conditions on the appearance
of individual outer-retinal bands.}
\subsection{OCT System Setup and System Calibration}
An in-house-built SS-OCT system was employed for all measurements.
The OCT operated at a 100 kHz sweep rate with a 110 nm tuning range centered at
1060 nm. \rev{The axial resolution was approximately $6~\mu$m in air, corresponding to
$4.35~\mu$m in retinal tissue after division by an assumed average retinal refractive index
$n_{\mathrm{ret}}=1.38$ at 1060~nm~\cite{Hariri2009AxialResolution}.} The lateral resolution was
$\sim 12.9~\mu$m. \rev{The experimental
B-scans were sampled laterally at $10~\mu$m per A-line.} The numerical
aperture (NA) of the sample objective scanning lens was 0.05.
The acquired data were stored as linear OCT intensity, and all experimental fitting
and attenuation reconstruction were performed in the linear-intensity domain.
Display transformations were applied only for image visualization and did not enter fitting or attenuation reconstruction.

To relate the sample-generated intensity in Eq.~\eqref{eq:oct_signal_relation}
to the acquired digital signal while accounting for system-dependent effects,
including confocal focusing and sensitivity roll-off, we write
\cite{VanLeeuwen2003AxialPSF,Faber2004Attenuation}
\begin{equation}
\label{eq:raw_oct_system}
\begin{aligned}
I_{\mathrm{acq}}(z)-I_{\mathrm{bg}}(z)
&\propto \beta H(z-z_f)T(z)I_{\mathrm{s}}(z) \\
&\propto \beta H(z-z_f)T(z)R(z)\mu(z)L(z),
\end{aligned}
\end{equation}
where $I_{\mathrm{bg}}(z)$ is the background signal, $T(z)$ denotes the sensitivity
fall-off, and $H(z-z_f)$ is the confocal
focusing function that describes the depth-dependent modulation of the OCT signal
away from the focal plane.
\rev{The depth-dependent confocal response was removed before fitting and attenuation
reconstruction; it was not absorbed into $C_{ab}$. After this correction, only residual
multiplicative scaling that is approximately constant within a local reconstruction interval
is included in $C_{ab}$ and absorbed into the fitted effective amplitude $A_{ab}$.}
Here $z_f$ is the \rev{focal-plane position in the acquisition-depth coordinate}. Owing to the long coherence
length of the light source ($\sim 12$~mm), the sensitivity roll-off is negligible within the 
imaging depth of approximately $3$~mm.

The confocal focusing function is modeled as
\cite{Faber2004Attenuation,Li2020RobustAttenuation}
\begin{equation}
\label{eq:confocal_function}
H(z - z_f) = 
\left[
\left(
\frac{z - z_f}{z_R}
\right)^2 + 1
\right]^{-1},
\end{equation}
where $z_R$ denotes the Rayleigh length of the incident Gaussian beam.

To estimate the system parameters, a planar mirror was imaged at two different depths
\cite{Dwork2016ConfocalFunction,Wang2021PSFCalibration},
producing two OCT signals denoted as \textit{image1} and \textit{image2}. Let $\Delta z$ be the 
difference in surface depth between the two measurements. For any depth $z$, the relation
\begin{equation}
\label{eq:mirror_relation}
\{ R(z)\mu(z)L(z) \}_{\mathrm{image1}}
=
\{ R(z+\Delta z)\mu(z+\Delta z)L(z+\Delta z) \}_{\mathrm{image2}}
\end{equation}
holds. Combining Eq.~\eqref{eq:raw_oct_system} with Eq.~\eqref{eq:mirror_relation} for the two images yields
\begin{equation}
\label{eq:calibration_ratio}
\left[
\frac{I_{\mathrm{acq}}(z)-I_{\mathrm{bg}}(z)}{H(z - z_f)}
\right]_{\mathrm{image1}}
=
\left[
\frac{I_{\mathrm{acq}}(z+\Delta z)-I_{\mathrm{bg}}(z+\Delta z)}
{H(z+\Delta z - z_f)}
\right]_{\mathrm{image2}}.
\end{equation}

The Rayleigh length $z_R$ and focal-plane position $z_f$ were determined by an exhaustive search that
minimized the mismatch between the two sides across all $z$. The corrected linear intensity
used for subsequent processing was
\begin{equation}
\label{eq:confocal_corrected_intensity}
\rev{I(z)\equiv I_{\mathrm{corr}}(z)
=\frac{I_{\mathrm{acq}}(z)-I_{\mathrm{bg}}(z)}{H(z-z_f)}.}
\end{equation}
\rev{The experimental B-scans used in Figs.~\ref{fig:phantom_maps},
\ref{fig:fig6_1}, and \ref{fig:bovine_first_order_assessment} were already outputs
of this upstream preprocessing. The same corrected linear intensity $I(z)$ was used to
obtain the within-scan fitting baselines, the conventional depth-resolved reconstruction,
$\hat{\mu}_0$, and $\hat{\mu}_1$. Consequently, $H(z-z_f)$ does not contribute an
additional depth-dependent term to the perturbation expansion.}
The calibration constant
$\beta L(0)$ was obtained from the integrated OCT signal of a strong reflector, yielding 
a value of $2.01\times 10^4$~A.U.$\cdot$mm.

\subsection{Experimental attenuation reconstruction and bovine first-order assessment}
\rev{To obtain experimental fitting-based attenuation coefficients, a fit
support \([z_{nf0},z_{nf1}]\subseteq[z_a,z_b]\) was defined within each segmented layer
from a representative laterally averaged A-line. For the phantom and bovine-retina data,
the proximal endpoint of the first two tissue intervals was moved inward by 10\% of the
segmented-layer thickness, and the distal endpoint of every interval was moved inward by
five axial samples to exclude the interface peak and its shoulder.
The 10\% proximal exclusion was a fixed relative rule used for the broader proximal
interface shoulder in the first two intervals, whereas the five-sample distal guard was
determined from the resolution, smoothing-support, and segmentation-uncertainty
considerations described below; neither value should be interpreted as a universal
system-independent constant.
The five-sample guard band corresponds to $13.1~\mu$m for the experimental sampling, or
approximately three
times the estimated $4.35~\mu$m axial resolution in retinal tissue. Operationally, the boundary
guard was chosen conservatively from the sum of the axial-resolution support, the
half-support of any axial smoothing, and a two-sample boundary-perturbation allowance
used as a practical proxy for segmentation uncertainty, not a measured segmentation
error. The final exponential fits used no axial smoothing, so these contributions were
two, zero, and two samples, respectively; one additional sample was retained as a conservative
safety margin, giving the five-sample guard.
Every retained support was also required to satisfy the minimum-support criterion stated
above.}
\rev{The within-scan exponential-fit baseline \(\mu_{\mathrm{fit}}\) was obtained by fitting the OCT intensity profile with an exponential model,}
\begin{equation}
\label{eq:fitting_reference}
I(z) \propto \mu_{\mathrm{fit}}\exp(-2\mu_{\mathrm{fit}}z), \qquad z\in(z_{nf0},z_{nf1}).
\end{equation}
\rev{This fit uses the same single-exponential signal model and OCT data as the
zeroth-order estimator in Eq.~\eqref{eq:zeroth_order_fit}. In the primary bovine
analysis, its layer-wise values supply the zeroth-order baseline. The fitted
curve in Fig.~\ref{fig:fig6_1}(d) therefore provides a within-scan baseline for
describing profile differences, not an independent test of reconstruction
accuracy or a measurement of intrinsic optical properties.}
The corresponding \rev{fitting-derived} backscattering fractions were estimated following the method described by Wang \emph{et al.}~\cite{Wang2023StationaryIteration}.

\rev{For the two-layer phantom reconstructions, representative A-lines were
averaged over 21 adjacent lateral positions to obtain the within-scan fitted
baseline. The upper and lower fitting supports contained 130 and 160 samples for phantom~1
and 110 and 178 samples for phantom~2; the shortest phantom support therefore contained
110 samples (a first-to-last span of $286.4~\mu$m), well above the 10-sample minimum for the experimental
sampling. Separately acquired scans of the same physical single-layer slabs,
recorded before assembly using the same OCT system and preprocessing, provided a
calibration benchmark \(\mu_{\mathrm{cal}}\) that did not reuse the assembled B-scans.
Three nonoverlapping lateral ROIs were examined per scan, and each ROI-averaged
linear-intensity profile was fitted to a single exponential after excluding
25 axial samples on either side of the detected slab boundaries. An ROI fit
was retained when its intensity-domain $R^2$ was at least 0.8; a scan required
at least two retained ROIs. The mean of the retained ROI fits defined each
accepted scan value, and the reported standard deviation was calculated across
accepted scans rather than treating ROIs as independent specimens. The accepted
scan counts were 3/6 and 4/8 for the top and bottom slabs of phantom~1 and
7/11 and 6/9 for those of phantom~2. In the same order, the separately acquired
\(\mu_{\mathrm{cal}}\) values were $0.657\pm0.053$, $0.812\pm0.150$,
$0.535\pm0.036$, and $0.765\pm0.092~\mathrm{mm}^{-1}$ (mean $\pm$ between-scan
standard deviation across accepted scans, not across independently fabricated
slabs). This separate-acquisition benchmark removes the
same-B-scan fitting dependence but remains an OCT-model-based effective
attenuation estimate, not an independent measurement of intrinsic optical
properties.}

For the primary bovine retinal B-scan shown in Fig.~\ref{fig:fig6_1}, the OCT intensity was averaged over 11 adjacent A-lines centered on the location indicated by the dashed line in Fig.~\ref{fig:fig6_1}(b). \rev{The resulting inner-retina, outer-retina/RPE, choroid, and sclera fitting supports contained 42, 33, 78, and 372 axial samples, respectively; the shortest support therefore comprised 33 samples (a first-to-last span of $84.1~\mu$m), well above the stated minimum. Shifting each endpoint independently by up to two samples, approximately one in-tissue axial-resolution element, produced attenuation-fit coefficients of variation of 7.7\%, 12.4\%, 6.7\%, and 1.2\% for the four regions. The largest variation, 12.4\%, occurred in the relatively thin outer-retina/RPE support, illustrating why a minimum-support criterion and an explicit interface guard are needed.} Within each \rev{retained fitting support}, the averaged intensity was fitted using Eq.~\eqref{eq:fitting_reference}. The resulting layer-wise attenuation and amplitude parameters were used as fixed zeroth-order baseline values within the corresponding segmented regions across the B-scan, while the first-order correction was calculated from the local linear OCT intensity in each A-line.

To isolate the local experimental information contributed by the first-order term, a second bovine retinal B-scan acquired within minutes of the primary scan was analyzed without supplying a vessel mask or vessel boundaries. \rev{The same four-region definitions used for the primary B-scan---inner retina, outer retina/RPE, choroid, and sclera---were applied to the second B-scan after surface alignment. The corresponding attenuation and amplitude parameters obtained from the primary scan were transferred without refitting when constructing \(\hat{\mu}_0\); no separate RPE-only fitting interval was introduced.} The first-order correction \(\hat{\mu}_1\) was calculated independently from the local linear OCT intensity in every A-line of the second scan. Two vessel-associated shadow centers were identified from minima in a laterally smoothed deep-tissue intensity score, yielding centers near A-lines 600 and 692. Each shadow ROI comprised the center A-line and five adjacent A-lines on either side, while A-lines 245--255 formed a remote control ROI. For the control-referenced profiles, each \(|\hat{\mu}_1|\) A-line was aligned to the local retinal surface and smoothed over seven axial samples. The median control profile was then subtracted from every shadow-ROI profile. The solid curves in Fig.~\ref{fig:bovine_first_order_assessment}(f) show the median differences, and the shaded regions show the 25th--75th percentile range across the 11 A-lines in each shadow ROI. \rev{For visualization only, the S1 and S2 arrows in panel (f) mark the dominant-magnitude extrema associated with the structurally inferred sites in panel (a); they do not represent independently validated vessel centers and were not used for ROI selection or reconstruction.}

\section{Results}

\subsection{Numerical Simulation: A-line Simulation}

\begin{figure}[!t]
\centering
\includegraphics[width=0.95\linewidth]{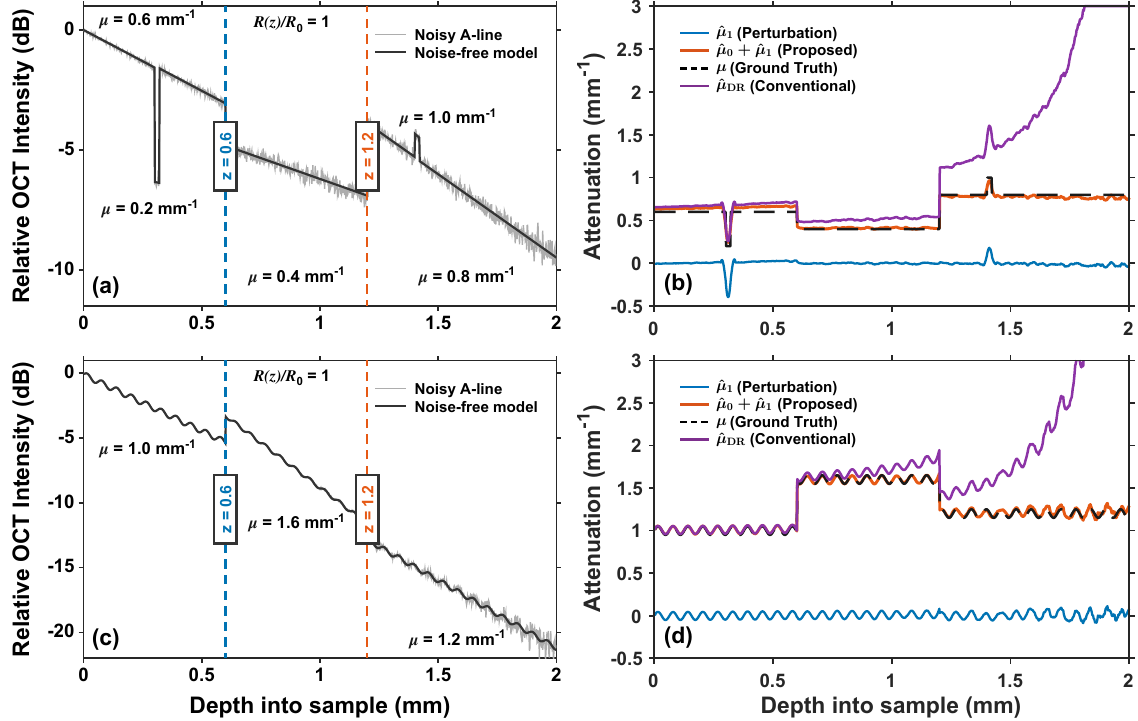}
\caption{\rev{Numerical A-line validation with incomplete signal decay.
(a,c) Simulated relative OCT intensity (dB) for step-like and smoothly varying
attenuation, respectively. Gray curves show noisy A-lines and black curves the
noise-free models; $R(z)/R_0=1$ in both cases. After unit-peak normalization,
the respective noise standard deviations were 0.01 and 0.001. Reconstruction
used linear intensity, not the dB display. (b,d) Corresponding attenuation
profiles from the conventional depth-resolved method and the proposed estimate
$\hat{\mu}_0+\hat{\mu}_1$.}}
\label{fig:fig1_1}
\end{figure}

Figure~\ref{fig:fig1_1} evaluates step-like and smoothly varying attenuation profiles under deliberate
finite-depth truncation.

As shown in Figs.~\ref{fig:fig1_1}(a) and \ref{fig:fig1_1}(c), the simulated OCT A-lines preserve the expected
depth-dependent decay and signal changes associated with the prescribed attenuation profiles.
\rev{The larger fluctuations in panel (a) arise from its tenfold higher noise standard
deviation relative to panel (c), with the difference accentuated by the logarithmic intensity
display; they do not indicate a different noise model or reconstruction-dependent effect.}
\rev{In the relative-dB plots, a constant-$\mu$ interval has slope $-20\mu/\ln 10$~dB/mm when $R(z)$ is constant: the $0.8~\mathrm{mm}^{-1}$ interval is therefore steeper than the $0.6~\mathrm{mm}^{-1}$ interval, and the $1.2~\mathrm{mm}^{-1}$ interval decays faster on average than the $1.0~\mathrm{mm}^{-1}$ interval. The narrow intensity changes at attenuation transitions arise from the multiplicative $\mu(z)$ factor in the signal model; the small sinusoidal modulation also changes the local slope.}
\rev{In both numerical cases,} the conventional depth-resolved method produced a depth-dependent increase
in the estimated attenuation. This bias was most evident near the distal end of the A-line, where
the missing signal tail violated the full-decay assumption. In comparison, the proposed estimate
followed the ground-truth attenuation profile more closely for both the step-like and smoothly
varying cases. This behavior is consistent with the first-order correction term, which reduces the
influence of finite-depth truncation without requiring integration over the unavailable distal tail.

\begin{figure}[!t]
\centering
\includegraphics[width=1\linewidth]{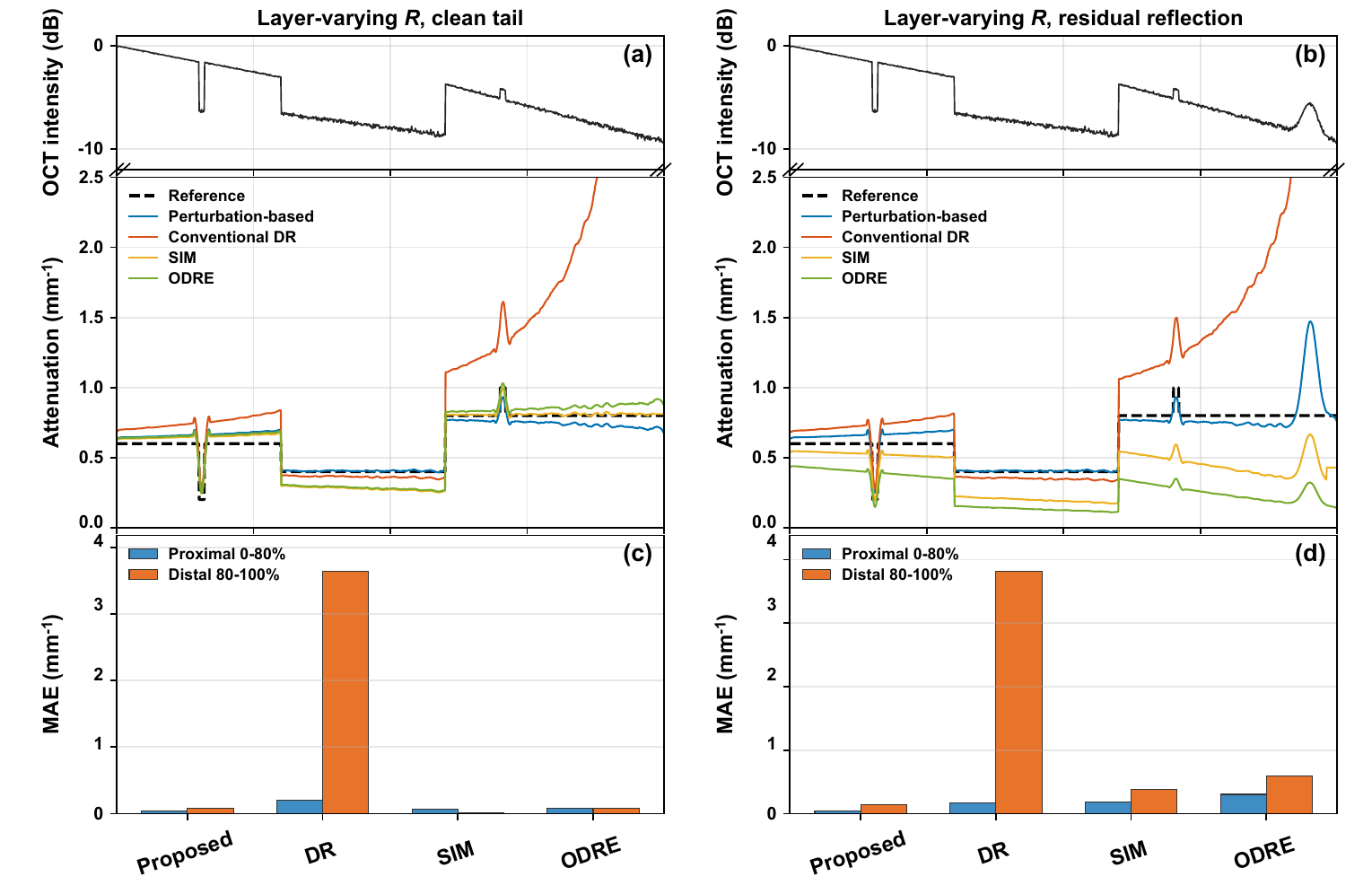}
\caption{\rev{Finite-tail attenuation benchmark under layer-varying backscattering.
(a,b) Simulated OCT intensity (top) and attenuation estimates (bottom) for the clean-tail
and residual-reflection cases. Curves show the prescribed reference, proposed method,
conventional depth-resolved reconstruction (DR), the independently implemented complete
stationary-iteration method (SIM), and optimized depth-resolved estimation (ODRE). (c,d) MAE
within masked samples in the proximal (0--80\%) and distal (80--100\%) portions of the
axial range.}}
\label{fig:r1c1_multimethod}
\end{figure}

\rev{Figure~\ref{fig:r1c1_multimethod} extends this comparison to recent reconstruction
approaches under layer-varying backscattering. In the clean-tail case, the full-mask MAEs
were $0.0435$, $0.0565$, and $0.0761~\mathrm{mm}^{-1}$ for the proposed method, SIM,
and ODRE, respectively, whereas conventional DR showed distal overestimation. With the
residual reflection, the corresponding MAEs were $0.0587$, $0.2245$, and
$0.3639~\mathrm{mm}^{-1}$. In the distal region, they were $0.1444$, $0.3757$, and
$0.5877~\mathrm{mm}^{-1}$, respectively. In both cases, SIM met its 1\% convergence
criterion after one joint update; the maximum absolute changes from its attenuation
initialization were $0.00146$ and $0.000645~\mathrm{mm}^{-1}$ for the clean-tail and
residual-reflection cases. Across the disclosed threshold and constraint sensitivity analysis,
the SIM attenuation changed by no more than $1.43\times10^{-4}~\mathrm{mm}^{-1}$ relative
to the primary setting. Its underestimation in Fig.~\ref{fig:r1c1_multimethod}(b) therefore
indicates that, under this implementation and benchmark, the complete joint iteration did
not remove sensitivity to contamination within the distal fitting interval.}

\subsection{\rev{Monte Carlo Evaluations}}
\subsubsection{\rev{Test 1: Multiple-Scattering Stress Test}}

\begin{figure}[htbp]
\centering
\includegraphics[width=1\linewidth]{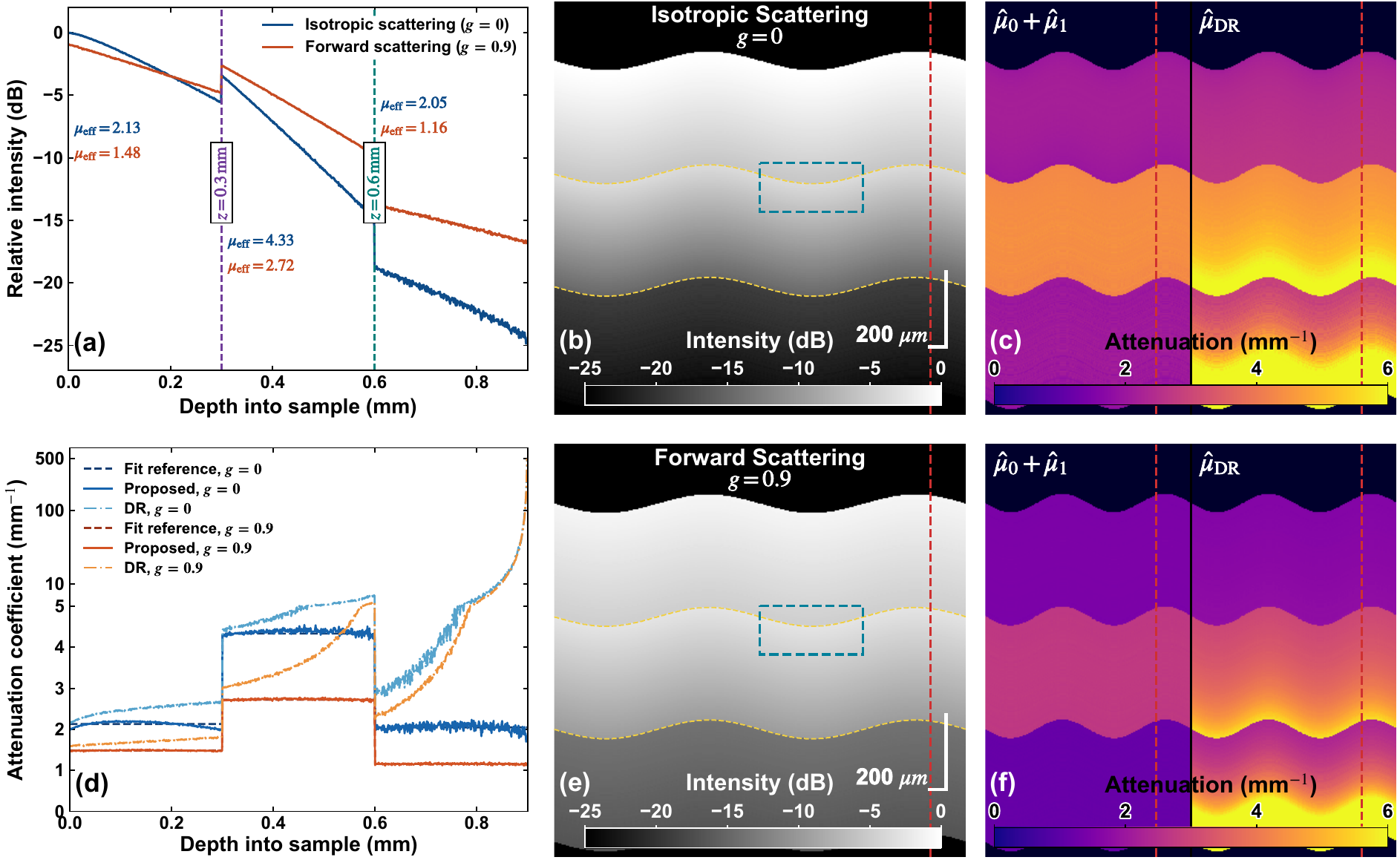}
\caption{\rev{Matched three-layer comparison at $g=0$ and $g=0.9$.
(a) Source A-lines on a shared relative-dB scale; dashed lines mark layer
boundaries, and colored labels give fitted effective attenuation
($\mathrm{mm}^{-1}$). (b,e) Identically translated displays of the respective
A-lines, not independent lateral simulations, on a shared $-25$ to 0~dB scale.
Yellow curves mark layer boundaries; cyan boxes highlight corresponding regions
around the first interface. (c,f) Proposed (left) and conventional depth-resolved
(right) reconstructions on a shared 0--$6~\mathrm{mm}^{-1}$ scale, saturated
above $6~\mathrm{mm}^{-1}$. Red dashed lines in (b,c,e,f) mark the representative
column. (d) Profiles and condition-specific fitted references, plotted linearly
through $5~\mathrm{mm}^{-1}$ and logarithmically above; the terminal conventional
rise is a finite-tail endpoint effect excluded from the central-80\% evaluation.}}
\label{fig:fig2_1}
\end{figure}

\rev{The first Monte Carlo test compared otherwise matched isotropic and strongly
forward-scattering three-layer profiles. Figure~\ref{fig:fig2_1}(a) shows that the
$g=0.9$ signal decayed more slowly and remained above the $g=0$ signal in the
distal layer.}

\rev{The fitted effective-attenuation references in layers 1--3 were, respectively,
$2.130$, $4.334$, and $2.049~\mathrm{mm}^{-1}$ for $g=0$, compared with
$1.484$, $2.721$, and $1.155~\mathrm{mm}^{-1}$ for $g=0.9$. In this matched
comparison, the $g=0.9$ signal had lower fitted effective attenuation than
the $g=0$ signal by $0.646$, $1.613$, and $0.894~\mathrm{mm}^{-1}$
(30.3\%, 37.2\%, and 43.6\%). The proposed $g=0.9$ estimates were likewise below
the $g=0$ fitted references by $0.642$, $1.599$, and $0.892~\mathrm{mm}^{-1}$.
This is a condition-dependent change in the detected signal, not a claim that
the method recovers intrinsic optical attenuation.}

\rev{Relative to the $g=0.9$ condition-matched fitted references, the proposed
layer-mean signed biases (reconstruction minus fit) were $+0.004$, $+0.014$,
and $+0.002~\mathrm{mm}^{-1}$, indicating slight overestimation of the fitted
effective levels. Its layer-wise MAEs were $0.008$, $0.018$, and
$0.013~\mathrm{mm}^{-1}$, whereas conventional tail integration gave $0.216$,
$0.989$, and $4.018~\mathrm{mm}^{-1}$.
Figure~\ref{fig:fig2_1}(d) shows that the proposed profiles followed the
condition-specific effective levels while the conventional profiles rose
toward the finite distal endpoint; Fig.~\ref{fig:fig2_1}(c,f) shows the
corresponding map differences. The benefit here is
reduced distal-error propagation, not reversal of the scattering-induced
downward shift or separation of photon scattering orders.}

\subsubsection{\rev{Test 2: Retina-Like Heterogeneous Reconstruction}}
\begin{figure}[htbp]
\centering
\includegraphics[width=1\linewidth]{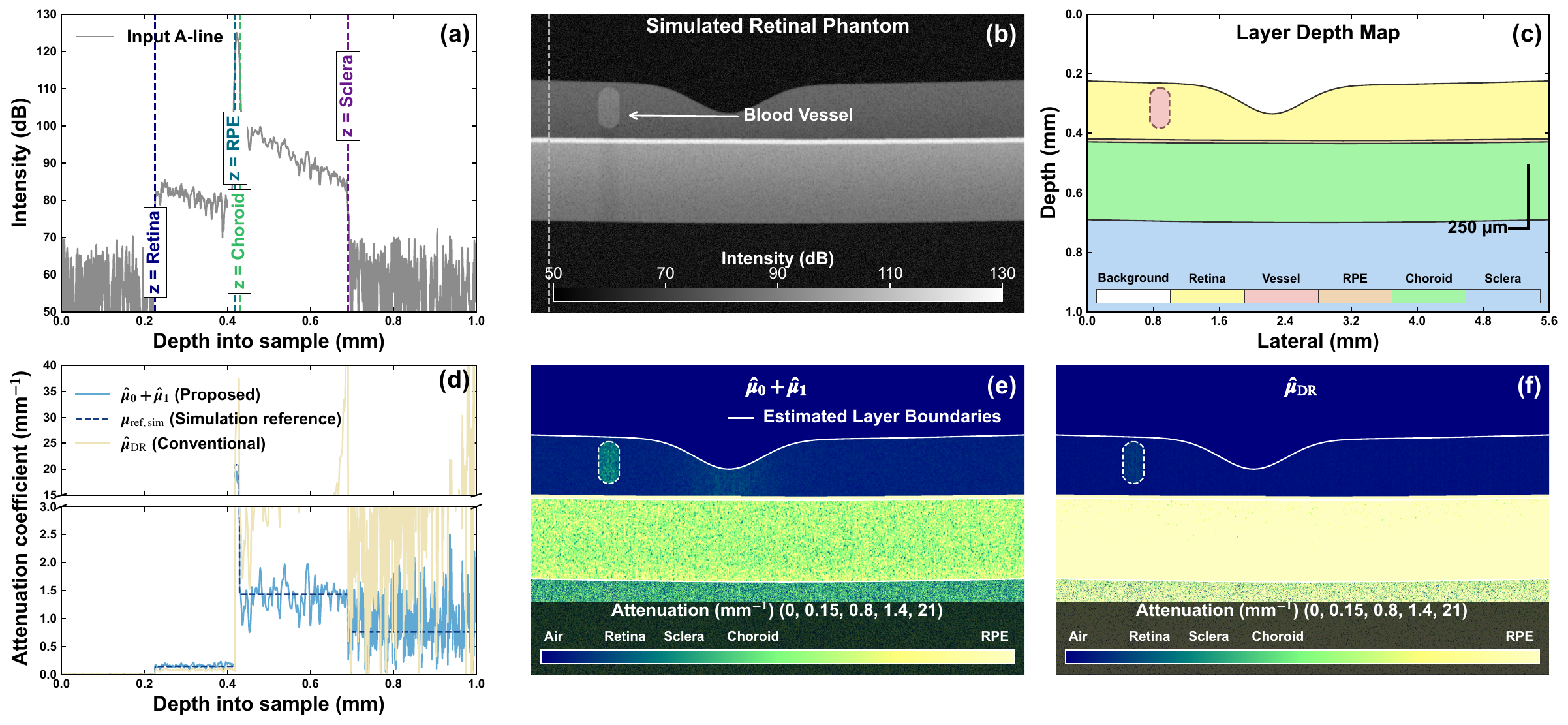}
\caption{\rev{Monte Carlo evaluation of a heterogeneous retina-like model with resolution-correlated Gaussian modulation. (a) Representative OCT A-line; (b) corresponding intensity image; (c) model geometry and tissue labels; (d) proposed and conventional attenuation profiles compared with the simulation-defined reference; (e,f) proposed and conventional attenuation maps, respectively. Values above the color bars in (e,f) are simulation-defined attenuation anchors used for nonlinear display mapping, not additional estimates. Panels (a,b,d--f) show the same stochastic realization.}}
\label{fig:fig3_1}
\end{figure}

\begin{figure}[htbp]
\centering
\includegraphics[width=1\linewidth]{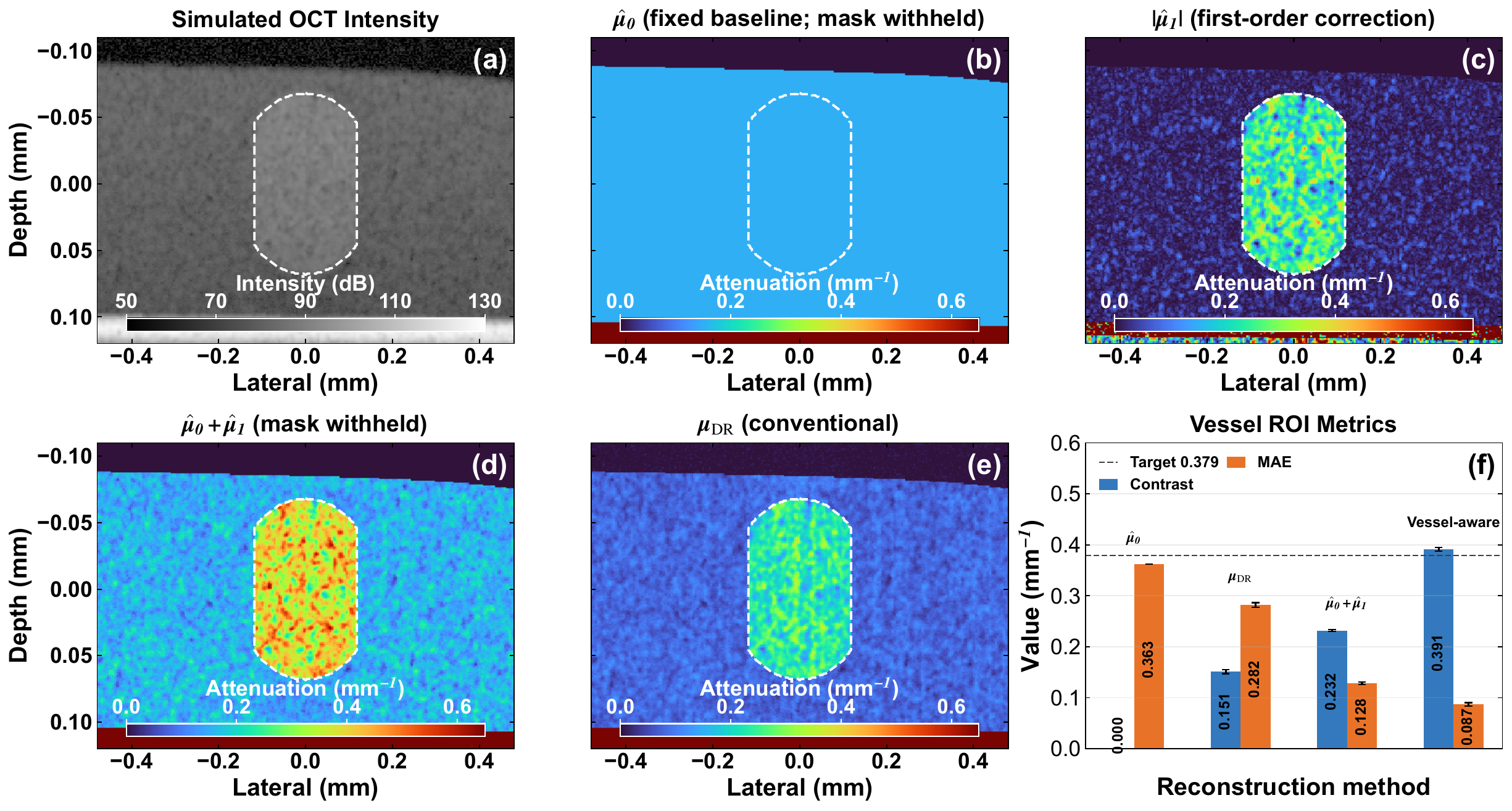}
\caption{\rev{Monte Carlo test of the first-order contribution with vessel segmentation withheld from reconstruction, using resolution-correlated Gaussian modulation. (a--e) Representative realization showing simulated OCT intensity, control-column-derived $\hat{\mu}_0$, $|\hat{\mu}_1|$, $\hat{\mu}_0+\hat{\mu}_1$, and $\hat{\mu}_{\mathrm{DR}}$, respectively. (f) Vessel contrast and vessel-ROI MAE across eight independent realizations (mean $\pm$ standard deviation); the dashed line indicates the segmentation-informed target contrast. ``Vessel-aware'' uses explicit vessel boundaries. Vessel contours in (a--e) are shown only for evaluation and were not used for parameter estimation or reconstruction.}}
\label{fig:vessel_ablation}
\end{figure}

\rev{The second Monte Carlo test evaluated the complete reconstruction in a curved,
heterogeneous retina-like geometry with an embedded vascular structure.}
Figure~\ref{fig:fig3_1}(a--c) shows the representative rendered A-line, OCT intensity image, and corresponding retinal model geometry. The region above the internal limiting membrane (ILM) represents air.

Due to the relatively low numerical aperture ($\mathrm{NA}=0.05$) and strong forward-scattering properties (high anisotropy factor $g$), multiply scattered photons contribute significantly to the detected signal. As a result, the simulation-defined reference attenuation is reduced, i.e., $\mu_{\mathrm{ref,sim}} < \mu_a + \mu_s$. The reconstructed attenuation profile and maps are shown in Fig.~\ref{fig:fig3_1}(d--f), along with comparisons to the conventional depth-resolved method and the proposed method. For visualization purposes, logarithmic compression and gamma correction were applied to enhance contrast in the attenuation maps.

The conventional method produced attenuation values above the simulation-defined
reference across much of the retinal model, with larger excursions near layer interfaces
and the maximum imaging depth. This behavior is consistent with the effect of
layer-dependent backscattering fraction and the inter-layer dependence introduced by tail
integration, as discussed in Section~2.1. Amplification of background noise and residual
deep signals further increased the distal estimates. The conventional reconstruction also
introduced artificial intra-layer variations in $\mu(z)$, particularly within the
choroid, which was homogeneous in the simulation model.

By comparison, the proposed method showed closer agreement with the simulation-defined
reference map and better preserved layer-wise attenuation contrast. \rev{These results were
obtained with the imposed Gaussian modulation and background floor; the proposed
reconstruction showed less propagation of distal-signal errors and more consistently
preserved the homogeneous layer structure.} Strongly scattering
regions, particularly in the sclera, were
more distinguishable in the attenuation map than in the corresponding intensity image in
Fig.~\ref{fig:fig3_1}(b). In addition, vessel-shadow artifacts visible in the intensity image were
reduced in the attenuation representation, suggesting improved structural contrast in the
simulated retinal model.

\subsubsection{\rev{Test 3: First-Order Contribution without Vessel Segmentation}}
\rev{The third Monte Carlo test directly isolates the information contributed by the first-order
correction under the resolution-correlated Gaussian model. Across eight independent
realizations, the fixed layer-wise zeroth-order estimate produced no vessel contrast, whereas
adding the local first-order term recovered $0.232\pm0.002~\mathrm{mm}^{-1}$, corresponding
to $61.1\pm0.6\%$ of the segmentation-informed target contrast of
$0.379~\mathrm{mm}^{-1}$. The vessel ROI MAE decreased from
$0.363~\mathrm{mm}^{-1}$ for $\hat{\mu}_0$ to
$0.128\pm0.002~\mathrm{mm}^{-1}$ for $\hat{\mu}_0+\hat{\mu}_1$, a reduction of
$64.7\pm0.6\%$. The conventional estimate $\hat{\mu}_{\mathrm{DR}}$ produced a contrast
of $0.151\pm0.004~\mathrm{mm}^{-1}$ and an MAE of
$0.282\pm0.004~\mathrm{mm}^{-1}$; thus, the corrected estimate obtained without vessel segmentation provided
approximately 1.5-fold greater vessel contrast and 55\% lower MAE. The vessel-aware
reconstruction remained more accurate, with a contrast of
$0.391\pm0.003~\mathrm{mm}^{-1}$ and an MAE of
$0.087\pm0.003~\mathrm{mm}^{-1}$, as expected because it used explicit vessel boundaries.
These results show that the first-order term restores substantial vessel-associated local
attenuation variation absent from a common layer-wise zeroth-order fit, although the residual
shadow response and lower accuracy relative to the vessel-aware reconstruction preclude
interpreting the result as complete recovery of intrinsic vessel attenuation.}

\subsection{Phantom Experiments}
\begin{figure}[!t]
\centering
\includegraphics[width=1\linewidth]{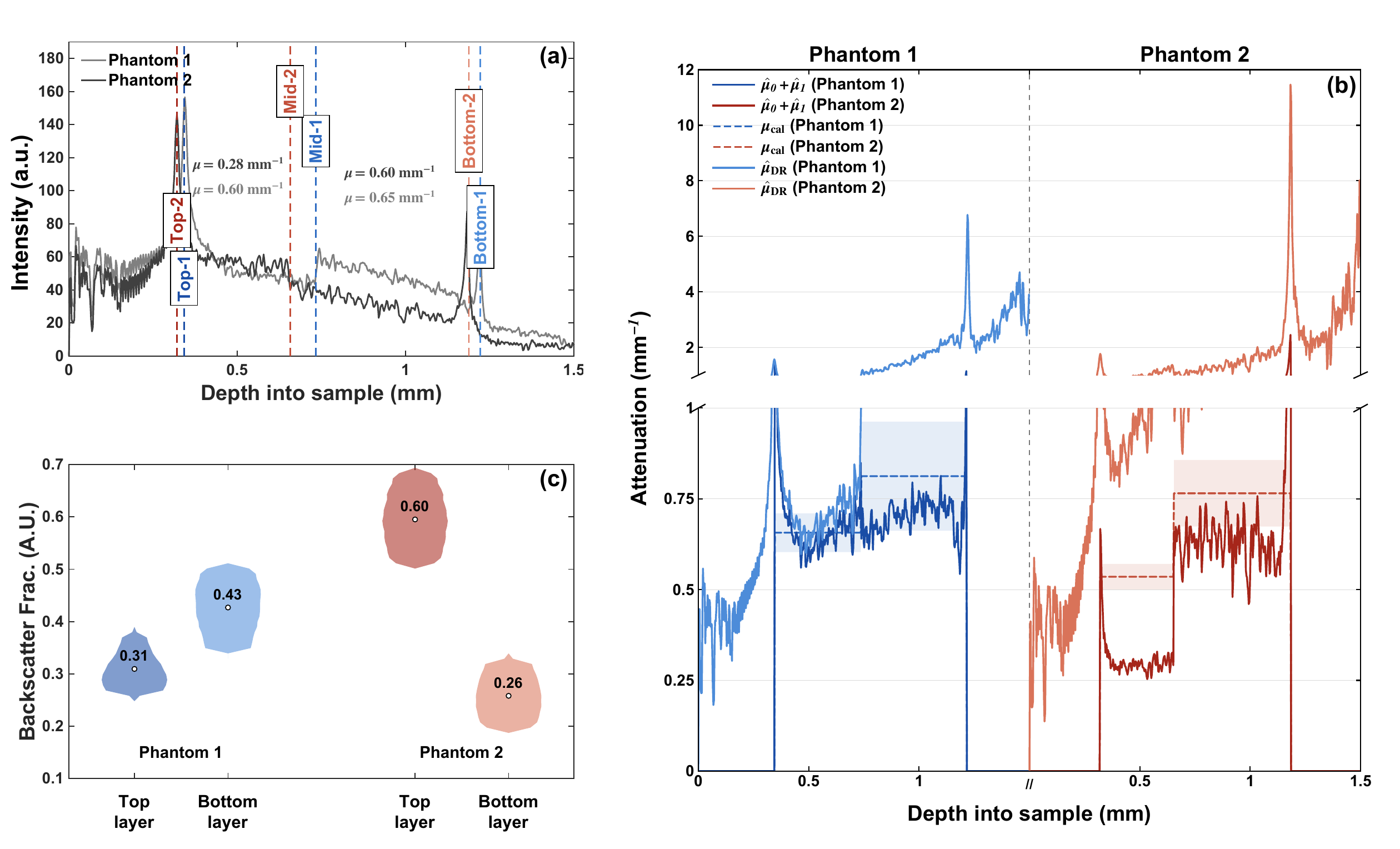}
\caption{\rev{Experimental profiles of two-layer silicone--dye--TiO$_2$ phantoms. (a) Representative linear-intensity OCT A-lines (a.u.) with annotated layer boundaries. Attenuation labels are within-scan fits used for reconstruction, not calibration values. (b) Proposed and conventional attenuation profiles for phantom~1 (left) and phantom~2 (right), compared with separately acquired single-layer calibration means $\mu_{\mathrm{cal}}$ (colored dashed lines) and between-scan standard deviations (shaded bands). The vertical dashed line separates the phantoms; the attenuation axis changes scale at $1~\mathrm{mm}^{-1}$ without omitting values. (c) Estimated layer-wise backscattering fractions.}}
\label{fig:phantom_profiles}
\end{figure}

Phantom~1 represented relatively weak interlayer optical-property contrast, whereas phantom~2
introduced a stronger layer-dependent mismatch. \rev{The estimated layer-wise backscattering
fractions in Fig.~\ref{fig:phantom_profiles}(c) show a larger top-to-bottom difference in
phantom~2. The OCT A-lines in Fig.~\ref{fig:phantom_profiles}(a) retain distal signal because
the finite phantom thickness and bottom air interface prevent complete decay within the
imaged depth.}

\begin{figure}[!t]
\centering
\includegraphics[width=1\linewidth]{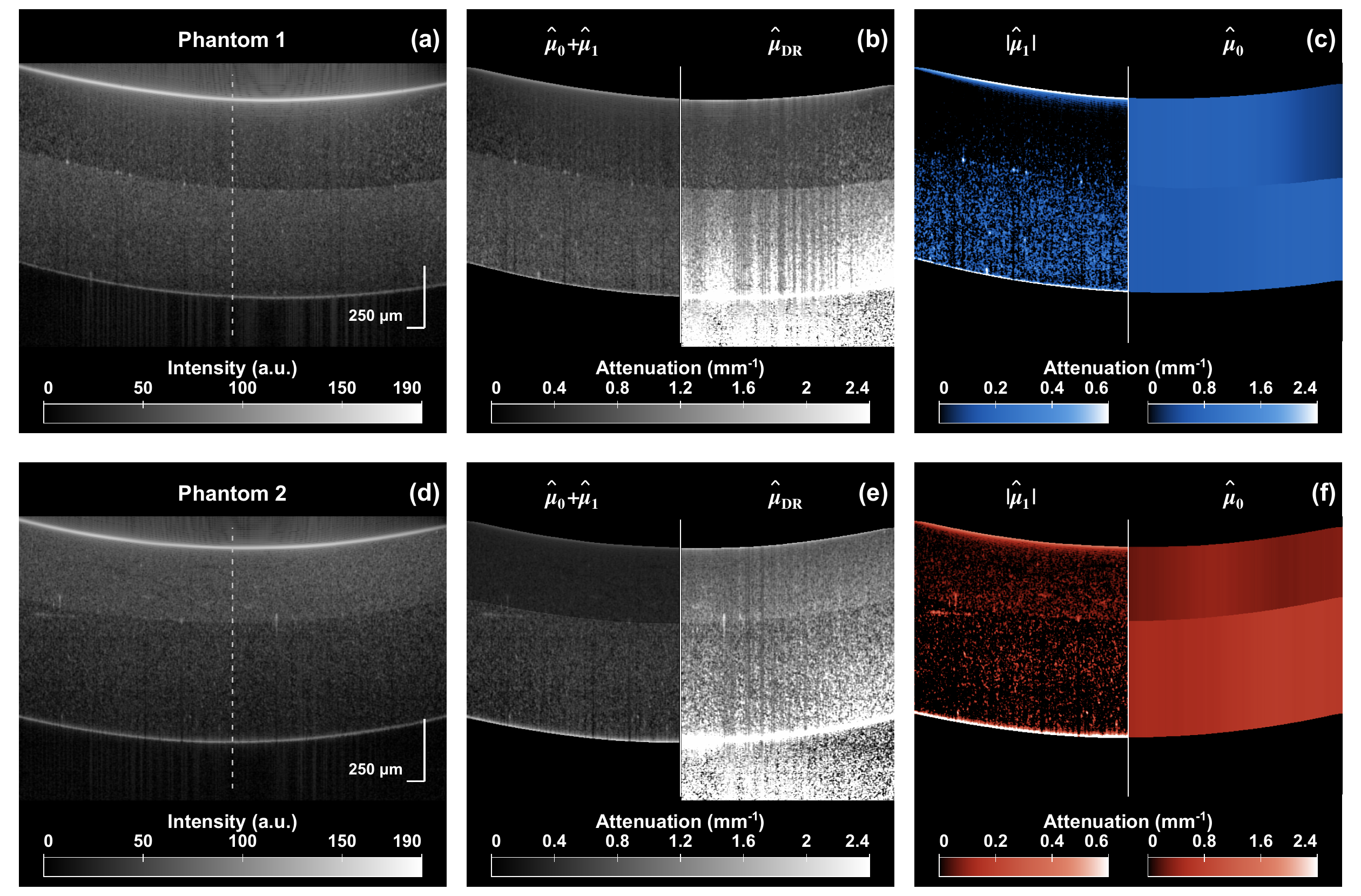}
\caption{\rev{B-scan comparison for the same phantoms. (a--c) Phantom~1; (d--f) phantom~2. (a,d) Recorded linear OCT intensity in arbitrary units on a shared 0--190 display scale. (b,e) Proposed $\hat{\mu}_0+\hat{\mu}_1$ on the left and conventional $\hat{\mu}_{\mathrm{DR}}$ on the right. (c,f) $|\hat{\mu}_1|$ on the left and $\hat{\mu}_0$ on the right. The solid central line in each paired map separates the two quantities, rather than indicating a tissue boundary. The $\hat{\mu}_0$ component images in (c,f) are laterally smoothed for display only; the final proposed maps in (b,e) use unsmoothed reconstruction components.}}
\label{fig:phantom_maps}
\end{figure}

\rev{In Fig.~\ref{fig:phantom_profiles}(b), the conventional estimate rose above the
separately acquired single-layer calibration values with depth, particularly near the
distal interfaces. Across the four layer supports in the representative profiles,
the proposed layer means were closer to the preassembly calibration values. The phantom~2
upper-layer proposed mean nevertheless remained below its calibration value
($0.293$ versus $0.535~\mathrm{mm}^{-1}$), which may reflect differences between
preassembly and assembled measurements as well as model limitations. The conventional upward drift is consistent with
layer-dependent backscattering and incomplete signal decay. The proposed maps also
show fewer vertical stripe artifacts than the conventional maps in
Fig.~\ref{fig:phantom_maps}(b,e). The component maps in
Fig.~\ref{fig:phantom_maps}(c,f) show that $\hat{\mu}_0$ establishes the dominant
layer-wise level, whereas $|\hat{\mu}_1|$ identifies local departures, particularly
near interfaces and spatially varying signal features.}

\subsection{Bovine Retina Experiments}
\begin{figure}[htb]
\centering
\includegraphics[width=1\linewidth]{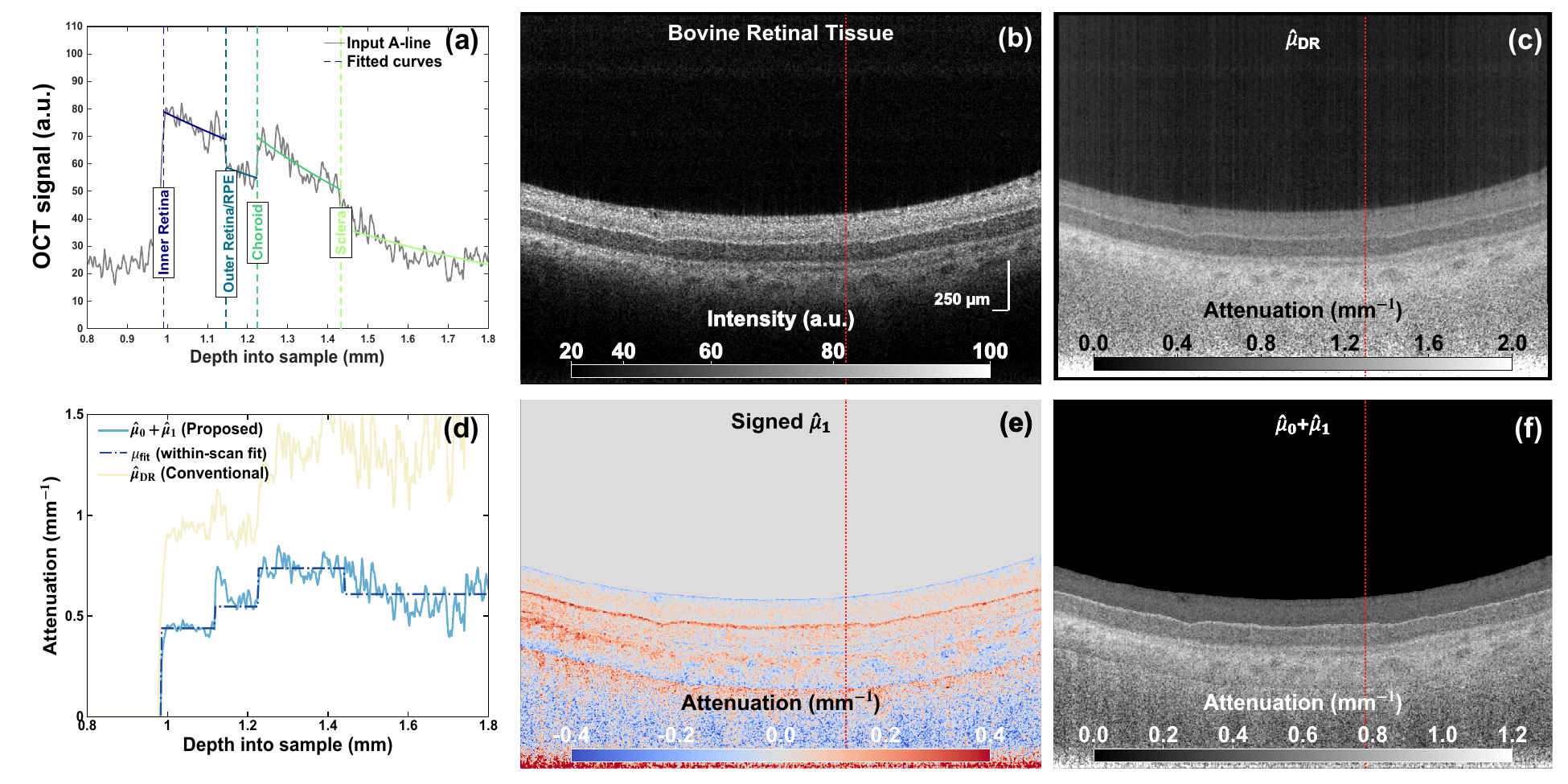}
\caption{Application to \emph{ex vivo} bovine retinal tissue. \rev{(a) Representative linear-intensity OCT profile in arbitrary units with piecewise exponential fits; the outer-retina/RPE label identifies the combined fitting interval. (b) B-scan displayed over 20--100 a.u.; squaring changes grayscale contrast only, while color-bar ticks and reconstruction use the original linear-intensity values. (c) Conventional estimate $\hat{\mu}_{\mathrm{DR}}$. (d) Attenuation profiles and the within-scan fit $\mu_{\mathrm{fit}}$, which supplies the layer-wise $\hat{\mu}_0$ but is not independent ground truth. (e) Signed correction $\hat{\mu}_1$. (f) Corrected estimate $\hat{\mu}_0+\hat{\mu}_1$. Red dashed lines in (b,c,e,f) mark the center of the 11-A-line window averaged for (a,d).}}
\label{fig:fig6_1}
\end{figure}

Figure~\ref{fig:fig6_1} compares conventional and perturbation-based reconstructions in an \emph{ex vivo}
bovine retina, together with representative A-line profiles.

\rev{In the \emph{ex vivo} bovine dataset, the individual bright photoreceptor/RPE bands
commonly identified in well-preserved \emph{in vivo} human retinal OCT were not clearly
separable. We therefore evaluated attenuation over a combined outer-retina/RPE region and did
not interpret the reconstruction as band-specific optical-property estimates.}

Consistent with the optical properties summarized in Table~1, the retina is dominated by highly forward-directed scattering and has a relatively small effective backscattering fraction. The more absorbing choroid may suppress the detected backscattered signal, whereas the highly scattering and less anisotropic sclera is expected to exhibit a larger effective backscattering fraction and stronger multiple-scattering contribution.

\rev{As shown in Fig.~\ref{fig:fig6_1}(c,d), the conventional depth-resolved profile varied substantially with depth and lay above the within-scan fitted baseline $\mu_{\mathrm{fit}}$ across the retinal and choroidal intervals. Its upward deviation became more pronounced in deeper tissue and near layer interfaces, consistent with depth-dependent backscattering and multiple-scattering contributions that violate the predominantly single-scattering, locally constant-backscattering approximation. Near the maximum imaging depth, the conventional profile rose sharply as background noise and residual deep signals entered its tail integral.}

\rev{The proposed profile did not show the same sharp distal rise in Fig.~\ref{fig:fig6_1}(d), consistent with its reduced dependence on the signal tail below the evaluated depth. The fitted curve $\mu_{\mathrm{fit}}$ is shown as the baseline used to form $\hat{\mu}_0$, not as an independent accuracy benchmark. The signed component map in Fig.~\ref{fig:fig6_1}(e) shows that \(\hat{\mu}_1\) introduces both positive and negative local deviations from this fixed layer-wise baseline. These refinements are retained in the direct sum \(\hat{\mu}_0+\hat{\mu}_1\) in Fig.~\ref{fig:fig6_1}(f), which maintained modest layer-wise lateral variability without the pronounced distal rise of the conventional profile. A component-wise decomposition and quantitative lateral-variability analysis are provided in Supplementary Fig.~S1.}

\begin{figure}[htb]
\centering
\includegraphics[width=1\linewidth]{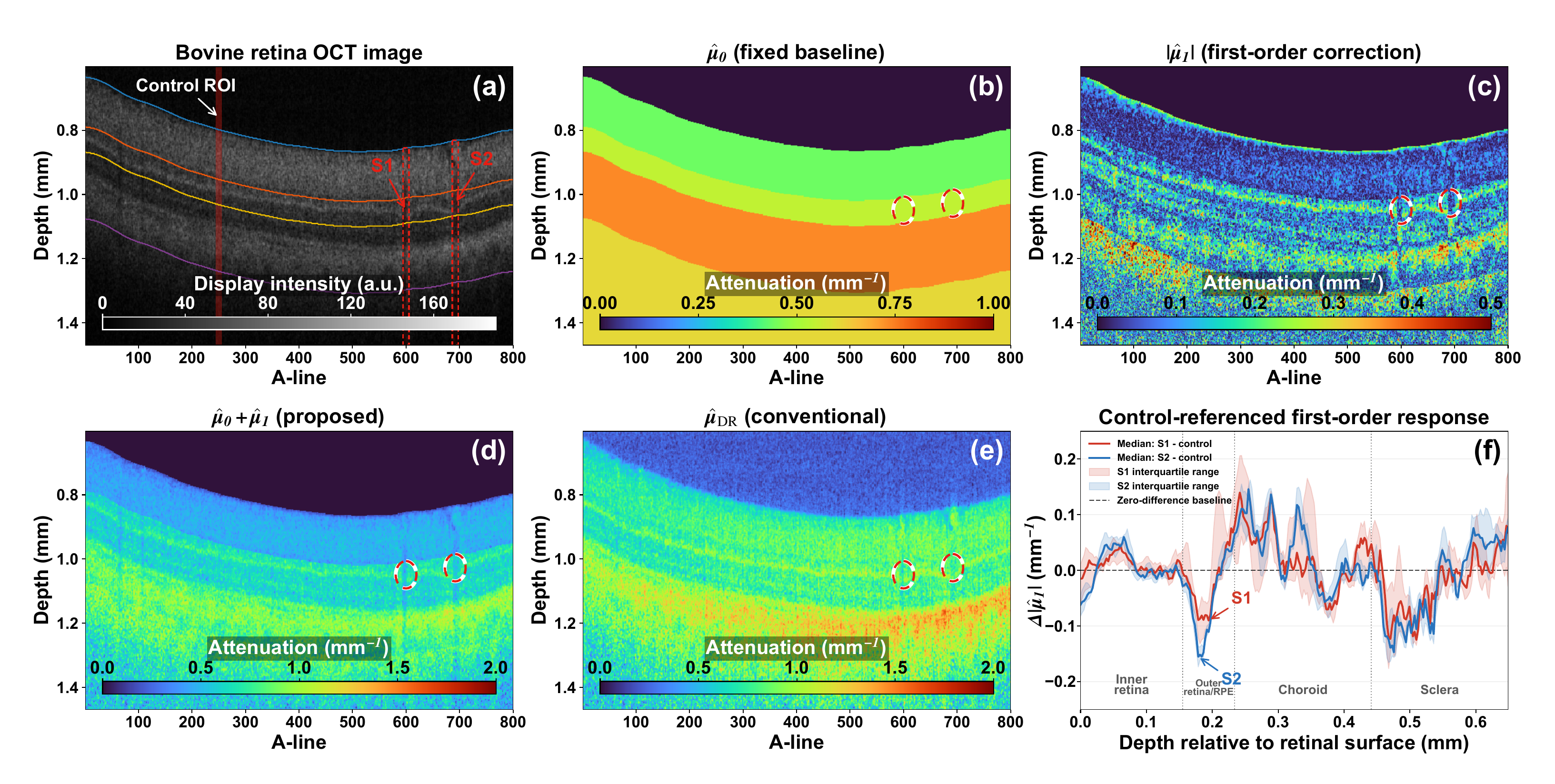}
\caption{\rev{Qualitative first-order assessment at two candidate vessel-associated
shadow sites in an \emph{ex vivo} bovine retina. (a) Structural OCT image with
segmented boundaries, the remote control ROI (translucent red band), and 11-A-line
S1/S2 sampling windows (dashed boxes). S1 and S2 are inferred from structural
shadows, not independently confirmed vessel centers. (b--e) Matching dashed ellipses
locate these sites in the fixed layer-wise $\hat{\mu}_0$ (0--1~mm$^{-1}$), correction
magnitude $|\hat{\mu}_1|$ (0--0.5~mm$^{-1}$), proposed $\hat{\mu}_0+\hat{\mu}_1$,
and conventional $\hat{\mu}_{\mathrm{DR}}$ maps, respectively; (d,e) share a
0--2~mm$^{-1}$ scale. (f) Surface-aligned, control-subtracted $|\hat{\mu}_1|$
profiles; curves and shading show ROI medians and interquartile ranges. The
horizontal dashed line marks zero difference, and vertical dotted lines delimit
the four tissue intervals.}}
\label{fig:bovine_first_order_assessment}
\end{figure}

\rev{Figure~\ref{fig:bovine_first_order_assessment} provides a qualitative experimental counterpart to the Monte Carlo first-order component test without vessel segmentation in a separate B-scan, where the structurally inferred shadow sites S1 and S2 are marked consistently across panels (a--e).} The fixed layer-wise \(\hat{\mu}_0\) in Fig.~\ref{fig:bovine_first_order_assessment}(b) contains no lateral response at \rev{these sites}, whereas \(|\hat{\mu}_1|\) in Fig.~\ref{fig:bovine_first_order_assessment}(c) exhibits localized depth-dependent responses at both positions. These refinements remain visible in the final \(\hat{\mu}_0+\hat{\mu}_1\) map in Fig.~\ref{fig:bovine_first_order_assessment}(d). \rev{With the layer parameters transferred from the primary scan, the combined outer-retina/RPE estimate remains within the shared 0--2~mm$^{-1}$ display range in panel (d). The component and final-map color scales are identified separately in the caption.} After surface alignment and subtraction of the remote control profile, both shadow ROIs show spatially coherent nonzero first-order responses over selected retinal depths in Fig.~\ref{fig:bovine_first_order_assessment}(f). \rev{As an a posteriori consistency check, the accumulated perturbation magnitude was below 0.020 for 95\% of pixels and remained below 0.033 throughout the combined outer-retina/RPE interval. These small values are consistent with using the first-order approximation for this dataset, but do not independently bound all neglected higher-order terms. The first-order term thus contributes a local response at the marked shadow sites that is absent from the fixed zeroth-order baseline. Because independent vessel localization and bovine vessel-attenuation measurements were unavailable, and the selected features include downstream shadow responses, this result is interpreted qualitatively rather than as recovery of intrinsic vessel attenuation.}

\section{Discussion}

The main contribution of this work is a perturbation-based attenuation reconstruction method for OCT measurements in which the distal A-line signal is unreliable or unavailable. Conventional depth-resolved methods compute attenuation using the signal below the evaluated depth; when this signal is truncated or contaminated by noise, residual reflections, or multiple scattering, the resulting tail integral can bias shallower estimates \cite{Vermeer2014DepthResolved}. Existing compensation strategies modify the distal residual-light or boundary term but remain within the tail-integral framework \cite{Smith2015AutomatedAttenuation,Li2020RobustAttenuation}. The proposed method instead separates a zeroth-order exponential estimate from a first-order correction term, thereby reducing dependence on the distal signal tail.

The zeroth-order term $\hat{\mu}_0$ is \rev{estimated from a retained fitting support and applied across the corresponding local reconstruction interval}, while the first-order correction $\hat{\mu}_1$ accounts for depth-dependent deviations using the signal accumulated between the upper \rev{reconstruction-interval} boundary and the current depth. This formulation avoids extrapolating or iteratively updating the unavailable distal tail. In the numerical A-line simulations, this benefit was most evident when the signal did not fully decay within the imaging range. \rev{In the matched three-layer Monte Carlo comparison, the proposed method tracked the condition-specific effective attenuation while reducing additional depth-dependent error from distal-signal propagation; it did not reverse the downward shift in effective attenuation between the $g=0$ and $g=0.9$ signals. Under the resolution-correlated Gaussian model, the first-order component test without vessel segmentation showed that the correction recovered approximately 61\% of the segmentation-informed vessel contrast and reduced vessel ROI MAE by approximately 65\% relative to the common layer-wise zeroth-order baseline, demonstrating that the correction restores substantial local attenuation variation without explicit vessel boundaries.}

In the silicone--dye--TiO$_2$ phantoms, the method reduced nonphysical stripe artifacts associated with bottom reflections and incomplete decay. The qualitative bovine assessment in Fig.~\ref{fig:bovine_first_order_assessment} provides an experimental counterpart: two vessel-associated shadow locations that were absent from a fixed layer-wise \(\hat{\mu}_0\) produced localized control-referenced responses in \(|\hat{\mu}_1|\), which were retained in the final reconstruction. \rev{In the bovine retina, it maintained modest layer-wise lateral variability without the pronounced distal rise seen in the conventional profile.} Together, these results indicate that the method is most useful when finite imaging depth and heterogeneous optical properties limit the reliability of tail-integral reconstruction.

\rev{The proposed method is complementary to the stationary-iteration method (SIM)
rather than a replacement for it \cite{Wang2023StationaryIteration}. SIM jointly estimates
depth-dependent attenuation and backscattering, whereas the present method reduces
sensitivity to unreliable distal-tail information. In the residual-reflection benchmark, the
complete independently implemented SIM update converged after one iteration and remained
close to its attenuation initialization, which underestimated the prescribed attenuation;
the proposed estimate remained closer to the reference. This result demonstrates the
tail-robustness distinction for the disclosed simulation and implementation, but it should
not be generalized beyond the evaluated fitting interval, constraints, or signal conditions.}

Several limitations remain. First, the method assumes that the backscattering fraction is locally constant within each \rev{local reconstruction interval}. This assumption is weaker than treating $R$ as constant over the full A-line, but it may still introduce bias near abrupt tissue boundaries or in regions with rapidly varying microstructure. \rev{Broad lateral averaging across such boundaries could introduce partial-volume bias. The 101- and 51-pixel operations reported above were display-only, not minimum reconstruction windows; in laterally heterogeneous tissue, optional stabilization should be omitted or restricted to segmentation-consistent, edge-aware neighborhoods selected relative to lateral resolution and feature scale.}

Second, in strongly multiple-scattering tissues, the reconstructed attenuation should be interpreted as an effective OCT attenuation parameter rather than an intrinsic microscopic scattering coefficient. Third, \rev{the phantom calibration values came from separate preassembly scans of the same slabs, whereas the bovine fitted baseline remains a within-scan fit; neither independently measures intrinsic optical properties.} The bovine shadow features in Fig.~\ref{fig:bovine_first_order_assessment} were also identified structurally, without angiographic or histological confirmation, and therefore support a local first-order response rather than quantitative vessel-attenuation recovery.

\rev{Fourth, the distinct bright photoreceptor/RPE bands typical of well-preserved \emph{in vivo} human retinal OCT were not resolved in the present \emph{ex vivo} bovine dataset. Possible contributors include species-specific anatomy, open-sky preparation, postmortem changes, tissue preservation, and acquisition conditions; their relative contributions cannot be distinguished without matched \emph{in vivo} imaging, histology, or an independent preservation assessment. This limits band-specific interpretation and direct extrapolation to human retina. The within-dataset comparison remains informative because all methods use the same acquired data and the experimental conclusion concerns distal-tail dependence, not individual outer-retinal bands. Validation of band-specific attenuation will require well-preserved specimens and \emph{in vivo} human data with independent anatomical references.}

Finally, the reconstruction remains sensitive to local fitting stability, speckle and background noise, and \rev{fitting-support selection}, making robust lateral processing and automated \rev{support selection} important for broader implementation.

Future work will focus on automated and reproducible selection of \rev{fitting supports}, cross-system validation, and evaluation across additional tissue types and disease models. Quantitative attenuation and backscattering measurements have already been explored for skin wound assessment, atherosclerotic plaque characterization, filtering bleb assessment, retinal nerve fiber layer integrity, and whole-brain imaging \cite{Ghosh2021SkinWound,Xu2008PlaqueBackscatter,VanSoest2010AtheroscleroticAttenuation,Kosugi2021FilteringBleb,Thepass2017Glaucoma,Lefebvre2017WholeMouseBrain}. Beyond diagnostic imaging, robust attenuation estimation may also be useful for OCT-related surgical guidance, including ophthalmic image-guided microsurgery, OCT-guided robotic intervention, and depth-aware tissue characterization in procedures such as deep anterior lamellar keratoplasty \cite{Lee2021CPOCTSubretinal,Yu2025TopologyDALK,Xiong2026RealtimeDALK}. The proposed framework may further support such computational and surgical-guidance workflows when finite-depth or heterogeneous OCT signals limit the reliability of conventional attenuation reconstruction \cite{Yu2026CornealMCML,Wang2026CornealPnP,Yu2026PhysicsBasedOCT,Guo2025Psi3D}.

\section{Conclusion}
This work presents a perturbation-based framework for reconstructing depth-resolved OCT attenuation with reduced dependence on the distal signal tail. By combining a zeroth-order exponential estimate with a first-order correction term, the method addresses a specific failure mode of conventional tail-integral reconstruction, in which unreliable distal signal information can bias attenuation estimates at shallower depths. \rev{Numerical A-line simulations showed reduced depth-dependent error against prescribed attenuation profiles compared with conventional depth-resolved reconstruction. Monte Carlo comparisons showed less distal drift against model-defined benchmarks. In the phantoms, representative profiles were closer to separately acquired single-layer OCT calibration values than conventional profiles, whereas the \emph{ex vivo} bovine comparison remained relative to a within-scan fitted baseline. Neither experimental comparison establishes intrinsic optical-property accuracy.} \rev{The Monte Carlo first-order component test without vessel segmentation} and qualitative bovine assessment further showed that the first-order correction restores local attenuation variation that is absent from a fixed layer-wise zeroth-order baseline. Together, these results demonstrate that the proposed framework provides a practical approach to depth-resolved reconstruction of effective OCT attenuation in heterogeneous samples when full A-line signal decay cannot be achieved.

\begin{backmatter}
\bmsection{Funding}
This work was supported by National Institutes of Health grant
1R01EY032127 (PI: Jin U. Kang).

\bmsection{Disclosures}
The authors declare no conflicts of interest.

\bmsection{Data Availability Statement}
Data underlying the results presented in this paper are available from the corresponding author upon reasonable request.
\rev{A MATLAB and Python reference implementation of the reconstruction method, with synthetic examples and tests, is available under the BSD 3-Clause License at \url{https://github.com/arcalibrorum/perturbation-oct-attenuation}. The repository does not contain the experimental data or reproduce every figure in this paper.}

\bigskip

\rev{See Supplement 1 for supporting content.}

\end{backmatter}


\bibliography{sample}






\clearpage
\setcounter{figure}{0}
\renewcommand{\figurename}{Supplementary Fig.}
\renewcommand{\thefigure}{S\arabic{figure}}

\begin{center}
{\Large\bfseries Supplementary Material}

\vspace{0.5em}
{\bfseries Perturbation-Based Reconstruction of Depth-Resolved OCT Attenuation without
Requiring a Fully Decayed Signal Tail}

\vspace{0.5em}
Jinglun Yu, Yaning Wang, Shuwen Wei, and Jin U. Kang
\end{center}

\begin{figure}[!ht]
\centering
\includegraphics[width=\linewidth]{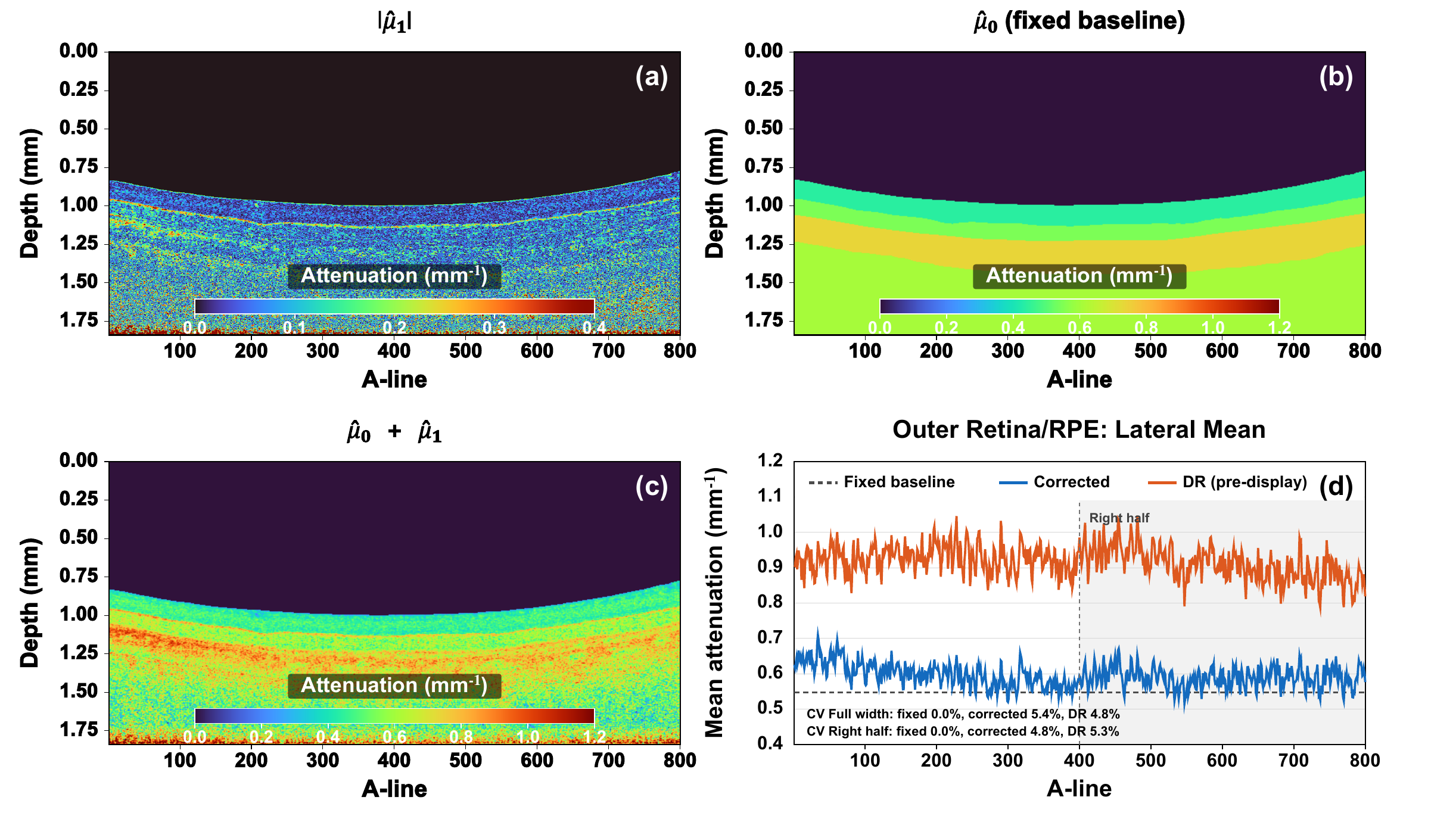}
\caption{Component and lateral-variability assessment for the primary
\emph{ex vivo} bovine retinal B-scan. (a) Magnitude of the first-order correction
$|\hat{\mu}_1|$ ($0$--$0.4~\mathrm{mm}^{-1}$). (b) Fixed layer-wise zeroth-order
baseline $\hat{\mu}_0$ and (c) direct sum $\hat{\mu}_0+\hat{\mu}_1$, both displayed
over $0$--$1.2~\mathrm{mm}^{-1}$. (d) Lateral mean attenuation within the combined
outer-retina/RPE interval for the fixed baseline, corrected estimate, and conventional
depth-resolved estimate before its separate per-column display offset. The same segmented
boundaries and five-sample interface guard were used for all three curves. The shaded
right half indicates A-lines 401--800. The displayed coefficients of variation (CVs)
use the sample standard deviation divided by the absolute lateral mean; the fixed
baseline has zero CV by construction.}
\label{fig:s1_lateral_uniformity}
\end{figure}

\end{document}